# Spintronic Sources of Ultrashort Terahertz Electromagnetic Pulses

T. S. Seifert[1*†], Liang Chen[2*†], Z. X. Wei[2], T. Kampfrath[1,3†], J. Qi[2†]

1 Department of Physics, Freie Universität Berlin, 14195 Berlin, Germany

2 State Key Laboratory of Electronic Thin Films and Integrated Devices, University of Electronic Science and Technology of China, Chengdu 610054, China

3 Fritz Haber Institute of the Max Planck Society, 14195 Berlin, Germany

* Authors contributed equally

†Email: tom.seifert@fu-berlin.de; chengliang@uestc.edu.cn; tobias.kampfrath@fu-berlin.de; jbqi@uestc.edu.cn

## Abstract

Spintronic terahertz emitters are novel, broadband and efficient sources of terahertz radiation, which emerged at the intersection of ultrafast spintronics and terahertz photonics. They are based on efficient spin-current generation, spin-to-charge-current conversion and current-to-field conversion at terahertz rates. In this Editorial, we review the recent developments and applications, the current understanding of the physical processes as well as the future challenges and perspectives of broadband spintronic terahertz emitters.

1. **Introduction**

1.1 *Terahertz time-domain spectroscopy*

Terahertz (THz) radiation covers the range from about 0.1 to 30 THz. It holds great promise for basic research and future applications [1, 2] because the THz frequency range coincides with many low-energy modes in all phases of matter, i.e., plasmas, gases, liquids and solids [3]. For example, THz radiation can resonantly couple to conduction-electron transport, plasmons, excitons, Cooper pairs, phonons or magnons [4]. Thus, THz spectroscopy is a powerful tool to study fundamental processes in a wide range of materials.

THz radiation not only serves as a probe: The development of high-amplitude THz sources enables the control of collective excitations of matter [5-7] such as magnons in magnets [8-11] or driving of phonons [12-16]. Currently, THz electric fields reach peak strengths of the order of 1 MV/cm in table-top systems, and they exceed ~10 MV/cm in large-scale user facilities such as free-electron lasers [17]. Upon excitation with intense THz pulses, ultrafast switching between different phases of matter (e.g., topological, magnetic and structural) was observed recently [8, 18-25]. THz excitation can also be combined with other well-established experimental probes, such as angle-resolved photoemission spectroscopy [26], scanning tunneling microscopy [27-29] or X-ray diffraction [30, 31]. Merging THz spectroscopy with such powerful probing techniques can yield unique multidimensional insights into fundamental processes on ultrafast time scales.

In terms of applications, THz imaging and sensing have recently gained considerable interest [32]. For instance, THz imaging holds great promise for biomedical and security applications since THz waves can be transmitted through living biological systems without harm, in contrast to wavelengths in the ultraviolet to X-ray range. THz imaging can be used for quality control and inspection such as in pharmaceutical [33], manufacturing [34, 35] and security applications [36]. Meanwhile, THz radar is also under intense development because of its potential for autonomous driving and military applications [37, 38].

The recent development of sub-wavelength THz microscopy, for instance by using sub-wavelength apertures [32, 39], tip-like probes or the near-field of the THz source [32, 40], led to spatial resolutions deep into the micrometer range or even down to tens of nanometers, i.e., far below the THz wavelengths of about 10 μm to 3 mm. With the combined advantages of high spatial resolution and low-frequency ultrafast optical response, THz near-field techniques are foreseen to become versatile tools to reveal the THz response of materials at the nanoscale.

Based on these promising prospects for THz radiation, the exploration and development of more capable THz sources with high efficiency, broad frequency bandwidth and easy access has become one of the most important goals in the THz-photonics community. Thus, we will focus on table-top laser-driven approaches for the generation of THz radiation in the following.

1.2 *Generation of THz radiation*

The generation of electromagnetic radiation, including THz radiation, requires a time-dependent charge-current density **J**. In the frequency domain, the electric field **E** of the resulting electromagnetic wave is determined by the wave equation [41]

$$-\nabla \times (\nabla \times \mathbf{E}) + \frac{n^2 \omega^2}{c^2} \mathbf{E} = -\frac{iZ_0 \omega}{c} \mathbf{J} \qquad (1).$$

Here, $\nabla$ is the spatial derivative (Nabla operator), $\omega/2\pi$ is the frequency, $n$ is the refractive index, $c$ is the vacuum speed of light, and $Z_0 \approx 377\ \Omega$ is the free-space impedance. In general, **E**, **J** and $n$ are complex-valued and depend on spatial position and frequency.

According to Eq. (1), two conditions must be fulfilled to obtain a broadband and high-amplitude THz emitter: (a) Maximum amplitude and bandwidth of **J**, and (b) constructive superposition, i.e., phase matching, of all THz waves generated throughout the volume of the emitter [42]. Condition (b) critically depends on the spatial distribution of the material-specific parameter $n$.

1.2.1  *Ultrafast photocurrents*

When the current **J** is dominated by the electrons of the emitter material, it has contributions from the orbital motion of the electrons and from the spinning current of their spins [43]. Accordingly, the total source current density can be written as a sum

$$\mathbf{J} = \nabla \times \mathbf{M} + \mathbf{J}_{\text{orb}} \qquad (2)$$

of terms due to (i) the electron spin density (magnetization) **M** and (ii) the orbital electron currents $\mathbf{J}_{\text{orb}}$ [41]. Both contributions can be driven resonantly and off-resonantly when a material is illuminated with light, as discussed in more detail in the following.

 (i) A time-dependent magnetization $\mathbf{M}(t)$ gives rise to the emission of magnetic-dipole radiation. An example is the THz emission that arises from the ultrafast magnetization quenching in a ferromagnet by resonant excitation of conduction electrons with a femtosecond laser pulse [44-46]. A nonresonantly driven term (i) is, for instance, possible by the inverse Faraday effect in an optically transparent material, in which torque is exerted on **M**, that is, magnon modes are excited by the driving laser pulse [47-49]. While mechanism (i) is highly interesting for the detection of magnetization dynamics, it typically lacks efficiency compared to other THz-generation mechanisms described by term (ii) because magnetic dipoles radiate less efficiently than electric dipoles.

(ii) In crystalline solids and for resonant optical excitation of electrons, $\mathbf{J}_{orb}$ can be further divided into shift and injection currents [50-53]. In a simplified picture, a shift current is a displacement of charges within the solid's unit cell. The current density typically has bipolar dynamics owing to the forward and subsequent backward flow of electronic charge density, convoluted with the pump-pulse intensity envelope. A typical example of a resonant shift current is the above-band-gap excitation of bulk GaAs [54] or the excitation of the surface-near regions of the topological insulator $Bi_2Se_3$ [53].

In contrast, injection currents arise from optically induced changes in the group velocity of electrons, and the current amplitude is proportional to a typically unipolar response function describing electronic relaxation, again convoluted with the pump-pulse intensity envelope. Injection currents can be spin-polarized and are maximized for circularly polarized driving light [52]. They are by far less popular for THz-pulse generation and were found by, for instance, resonant excitation of electrons with circularly polarized light in Wurtzite semiconductors [52] or in topological insulators [55].

For below-band gap ultrafast optical excitation of transparent nonlinear optical crystals such as ZnTe, GaP and GaSe [1, 2], the photocurrent generation mechanism is referred to as optical rectification. This off-resonant process can be understood as a shift current with vanishing relaxation time.

### 1.1.1 *Emission of the THz pulse*

The THz-generation efficiency of nonlinear optical crystals typically increases with thickness because a larger volume emits THz waves. However, an increased thickness often lowers the generated THz bandwidth due to phase-mismatch of the pump field and the emitted THz field and because of THz-radiation absorption by infrared-active phonons, thus violating condition (b). For instance, the Reststrahlen band in polar semiconducting crystals typically strongly attenuates emission of THz radiation in the range between about 5 and 10 THz. Thus, broad bandwidth and high efficiency are usually difficult to achieve simultaneously.

In general, the emitted THz-field amplitude is only sizeable if the ultrafast photocurrent exhibits a nonvanishing transient electric dipole moment and, thus, broken inversion symmetry. While in materials such as GaAs, ZnTe or GaP, inversion asymmetry is intrinsically provided by the crystal structure, it may also be induced by external means. An example is a photoconductive antenna, in which inversion symmetry is broken by applying a static electric field, which accelerates the photo-generated charge carriers. The resulting transient charge current emits a THz pulse [1, 2]. Photoconductive antennas are currently among the most widely used THz emitters due to their high efficiency and excellent usability. Typically, their bandwidth is limited to 0.1-5 THz [detrimental to condition (a)] by the relatively long-lived current flow in the used semiconductor materials (i.e., LT-GaAs) [56-59].

However, a notable exception is Ge, which is a nonpolar semiconductor and, therefore, lacks one-phonon absorption. Thus, Ge photoconductive THz emitters allowed one to extend the THz bandwidth $\Delta\omega_{\text{THz}}^{10\%}/2\pi$ up to about 10 THz, yet with reduced THz-generation efficiency [60]. Note that $\Delta\omega_{\text{THz}}^{10\%}$ is defined as the full width at 10% of the maximal THz electric field amplitude in the spectral domain at the detector position.

Another approach to generate THz radiation relies on laser-induced plasmas, in which the driving laser fields break the inversion symmetry [61]. Such THz plasma sources [62] can reach 0.1 mJ pulse energy at a large bandwidth of up to about 100 THz or up to 50 mJ pulse energy at 1 THz bandwidth. However, their energy fluctuation is typically larger than that of other commonly used THz sources due to the high degree of nonlinearity of the generation process. The large pump-pulse energies require costly amplified laser systems.

Therefore, new THz-emitter concepts need to be explored. They should simultaneously satisfy the requirements of high THz-generation efficiency, broad bandwidth, stability and easy access.

1.2 *Spintronic THz emitters*

The recent development of ultrafast spintronics and femtomagnetism paved the way toward a novel THz-emitter concept: magnetic heterostructures FM|NM based on thin films of a ferromagnetic metal (FM) and a normal metal (NM) (see Fig. 1) [63, 64]. After optical excitation of such a FM|NM stack, an ultrafast spin current flows across the interface from the in-plane magnetized FM into an adjacent NM layer. The ultrafast out-of-plane spin current is converted into an in-plane charge current through a spin-to-charge-current conversion (S2C) process. If the NM consists of a heavy metal like Pt, S2C predominantly happens in the NM layer by the inverse spin Hall effect (ISHE) [63-66]. The resulting sub-picosecond transient charge current emits electromagnetic waves with THz frequencies.

STEs have interesting properties:

1. They were shown to emit radiation with frequencies between 0.1 and 30 THz without any spectral gaps, thus outperforming photoconductive switches or nonlinear optical crystals (e.g., ZnTe and GaP) in terms of bandwidth [64].

2. STEs can be driven by virtually any pump wavelength [67, 68], in stark contrast to THz sources based on semiconductors.

3. The efficiency and long-term stability of STEs are also comparable with or even better than commonly used THz sources (see Fig. 7).
4. The fabrication procedure is based on well-established thin-film growth processes, which enable high-quality emitters with an outstanding homogeneity [64, 69].
5. The emitted THz field of the STE scales linearly with the excitation fluence up to about 0.1 mJ/cm$^2$ and only at about 5 mJ/cm$^2$ a permanent STE damage is induced [70].
6. One can conveniently grow large-area STEs to enable THz-high-field applications. In fact, by expanding the pump-laser beam to avoid optical damage of the STE, high pump powers of 5 W at 1 kHz repetition rate were used to generate THz peak electric fields of ~300 kV/cm [69] in the diffraction-limited focus of a Gaussian beam, and fields of 1 MV/cm are expected to be reached soon.
7. As a STE is made of thin metallic films, it can be easily microstructured, allowing for photonic nano-engineering to tailor the STE performance [65, 71-73].
8. The emitted THz pulse is linearly polarized, and the THz electric field is to a very good approximation perpendicular to the FM magnetization, which can be conveniently controlled by an external magnetic field [74], thereby enabling modulation of the THz emission at kHz rates [75].
9. Employing more complex magnetization patterns inside the STE allows one to create exotic THz-beam polarization states [76].
10. As the STE is only a few nanometers thick, no phase-matching conditions need to be fulfilled, and the substrate can act as an efficient heat sink. Moreover, the pump beam profile is preserved during the THz generation mechanism, making tight focusing possible.
11. The STE enables near-field imaging approaches with deep-subwavelength resolution [77].

Thus, the STE combines ultrabroad bandwidth, high efficiency, ease of use and flexibility in terms of geometry and design [64, 66]. These benefits demonstrate the potential of STEs in terms of THz applications and as a THz source even beyond the traditional THz community, which might be further fostered by the recent commercial availability of the STE (TeraSpinTec GmbH) [69, 78, 79].

In the past, several reviews were published that highlight specific aspects of spintronic THz emission and optically generated ultrafast currents, which we would like to recommend to the reader [80-86]. **In this Editorial, we review the fundamentals of STEs, various THz generation mechanisms, their optimization, applications, challenges and future perspectives. Special attention lies on the microscopic mechanisms and the resulting potential in terms of optimizing spintronic THz sources.**

2. **Key empirical results obtained from STEs**

STEs take advantage of ultrafast spin dynamics in FM|NM heterostructures to reach a high THz-generation efficiency and broad bandwidth. They complement the capabilities of traditional THz emitters with their spintronic principle of operation. In this section, the key empirical results for different STE realizations are presented.

2.1 *Model of the STE operation*

A typical STE is a heterostructure containing FM and NM layers with thicknesses of several nanometers. When excited with a femtosecond laser pulse, the structure emits THz radiation. We model this phenomenon as a result of four elementary processes (see Fig. 1):

(1) absorption of the optical pump pulse,

(2) generation of a spin current from the FM to the NM layer,

(3) conversion of the spin current into a transverse charge current, which

(4) acts as a source of a THz electromagnetic pulse.

In the following, we address processes (1)-(4) from a phenomenological viewpoint, whereas the microscopic mechanisms of processes (1)-(3) will be discussed in more detail in Section 3.

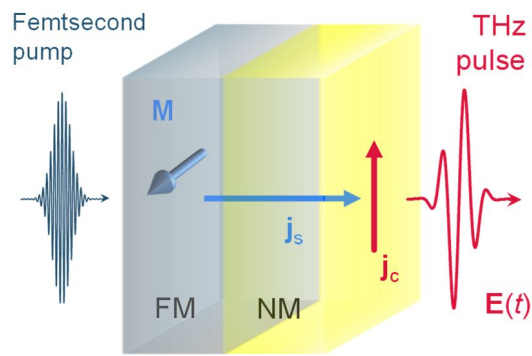

Figure 1. Electromagnetic-wave generation process in a spintronic THz emitter. An ultrashort laser pulse triggers a spin-current injection from the in-plane magnetized ferromagnetic metal (FM) into the normal metal (NM). In the next step, the spin-orbit interaction in the NM converts the spin current $\mathbf{j}_s$ into an in-plane charge current $\mathbf{j}_c$, which finally emits a THz electromagnetic

pulse. Directly behind the sample, the electric field of the linearly polarized THz pulse is perpendicular to the sample magnetization **M**.

(1) The absorbed pump-pulse energy is initially deposited primarily in the electronic system of the FM|NM stack. The resulting transient electron distributions are expected to be significantly different for the FM and NM layers.

(2) Macroscopically, a net spin transport from the FM to the NM layer is allowed because the FM|NM stack exhibits a broken inversion symmetry. Microscopically, the spin current can be carried by spin-polarized conduction electrons and/or magnons. Phenomenologically, spin currents can be driven by gradients of electrostatic potential, electron temperature and the spin accumulation [87]. If the FM is insulating and not excited by the pump pulse, the transient temperature gradient between the cold FM and the hot NM layer is the dominant driver of a magnon-type spin current through the interfacial spin Seebeck effect [88].

In contrast, a metallic FM layer such as Fe, Co or Ni exhibits a higher temperature following the pump absorption and, thus, aims at reducing its magnetization. The spin angular momentum can be released to the FM crystal lattice or by transport to the NM layer. In the framework of the Stoner model of ferromagnetism, one can show that the electron spin current arises from a gradient of the electronic temperature and the spin voltage between FM and NM layers [89]. The spin voltage $\mu_s$, also known as spin accumulation, can be understood as an excess of spin density (magnetization) in the FM layer that needs to be released to attain equilibrium. For a more detailed discussion, please see Sect. 3.1.

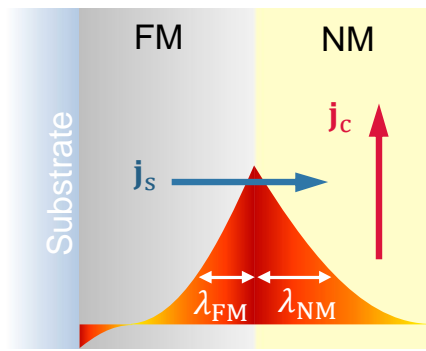

Figure 2. Ultrashort spin current following optical excitation in FM|NM stacks. The spin current decays over a length $\lambda_{\mathrm{NM}}$ inside the NM layer, which depends on the material and on frequency. In the FM layer, an analogous relaxation length $\lambda_{\mathrm{FM}}$ quantifies the mean FM layer depth, from which spin-polarized FM electrons still reach the FM|NM interface. Gueckstock, O., L. Nadvornik, M. Gradhand, et al., Adv. Mater., 33, p. e2006281, 2021; licensed under a Creative Commons Attribution (CC BY) license.

The resulting spin current with density $\mathbf{j}_s(z,t)$ flows parallel to the normal of the FM|NM stack (Fig. 2). Upon crossing the FM/NM interface, the spin current undergoes spin loss and reflection, and only a fraction of the spins is injected into the NM layer [64, 90, 91]. Inside the NM layer, the spin-current density $\mathbf{j}_s(z,t)$ decays with increasing depth $z > 0$ due to various electron scattering processes [63]. As shown in Fig. 2, the spatial decay is characterized by a relaxation length $\lambda_{\mathrm{NM}}$, which depends on the material and on frequency [92]. In the FM layer, an analogous relaxation length $\lambda_{\mathrm{FM}}$ exists. It quantifies the mean propagation length of spin-polarized FM electrons that still reach the FM/NM interface.

(3) While flowing, the spin current $\mathbf{j}_s = j_s \mathbf{u}_z$ is partially converted into a transient charge current $\mathbf{j}_c = j_c \mathbf{u}_z \times \mathbf{M}/|\mathbf{M}|$ by S2C. The direction of $\mathbf{j}_c$ is perpendicular to the sample normal and the FM magnetization. Phenomenologically, S2C can in the frequency domain be described by

$$j_c(z,\omega) = \gamma(z,\omega) j_s(z,\omega), \qquad (3a)$$

where the unitless parameter $\gamma$ quantifies the S2C strength. Both $j_s$ and $j_c$ have the unit s$^{-1}$ m$^{-2}$.

S2C arises from spin-orbit coupling, and two major mechanisms are considered: the ISHE and the inverse Rashba Edelstein effect (IREE). For the ISHE, $\gamma$ is accordingly called spin Hall angle. While the ISHE is symmetry-allowed in the bulk of any material, the IREE requires a locally broken inversion symmetry. Therefore, the IREE typically contributes to $\gamma(z,\omega)$ only at positions $z$ close to interfaces, such as of the FM and NM layers. Furthermore, the ISHE can often be considered instantaneous over the relevant frequency range 0-40 THz [79], implying a negligible frequency dependence of $\gamma$. In contrast, the IREE scales with the accumulated spin density and may, therefore, exhibit memory effects, which imply a frequency dependence of $\gamma$.

(4) In the final step, the ultrafast in-plane $\mathbf{j}_c$ emits THz radiation according to Eqs. (1) and (2). For FM|NM stack thicknesses much smaller than the THz wavelength and the THz attenuation length inside the FM and NM layers, the THz electric field is in the frequency domain approximately equal to [93]

$$\mathbf{E}(\omega) = eZ(\omega) \int \mathrm{d}z\, \mathbf{j}_c(z,\omega). \qquad (3b)$$

Here, $-e$ is the electron elementary charge, and $Z$ is the frequency-dependent impedance of the STE, which quantifies the current-to-electric-field conversion. It can be calculated according to

$$Z(\omega) = \frac{Z_0}{n_1(\omega) + n_2(\omega) + Z_0 G(\omega)}, \qquad (3c)$$

with the free-space impedance $Z_0 \approx 377\,\Omega$ and the refractive indices $n_1(\omega) \approx 1$ and $n_2(\omega)$ of air and the substrate, respectively. The THz sheet conductance $G(\omega)$ of the metal stack is $\int_0^{d_{\mathrm{NM}}+d_{\mathrm{FM}}} \mathrm{d}z\,\sigma(z,\omega)$, where $\sigma$ is the local THz conductivity of the metal layers.

## 2.2 *Experimental implementation*

*STE fabrication.* Because typical STEs consist of metallic films that are only a few nanometers thick, a variety of well-established thin-film deposition techniques can be used to grow STEs. The most prominent one is sputter deposition, which typically results in polycrystalline metallic films [64-66]. STEs made of single-crystalline FM and NM layers can be obtained via post-growth thermal annealing [94] or growth by molecular beam epitaxy [64, 95].

*Optical/THz setup.* The STE is usually driven by a pulsed femtosecond laser and used in combination with a THz time-domain spectroscopy (THz-TDS) system. This spectroscopic mode allows one to retrieve not only the amplitude but also the phase information of the THz pulse directly in the time domain [96]. To this end, the THz pulse is measured in a phase-resolved manner by electrooptic (EO) detection in a nonlinear optical crystal such as GaP or ZnTe. In these crystals, the THz electric field induces a transient birefringence via the linear EO effect. The refractive index of the EO crystal changes in proportion to the instantaneous THz electric field, which is sampled by a time-delayed optical probing pulse. By measuring the acquired probe-pulse ellipticity vs the delay between the THz and the probe pulse, the entire THz waveform is mapped out. A typical THz-TDS setup including the STE is schematically shown in Fig. 3.

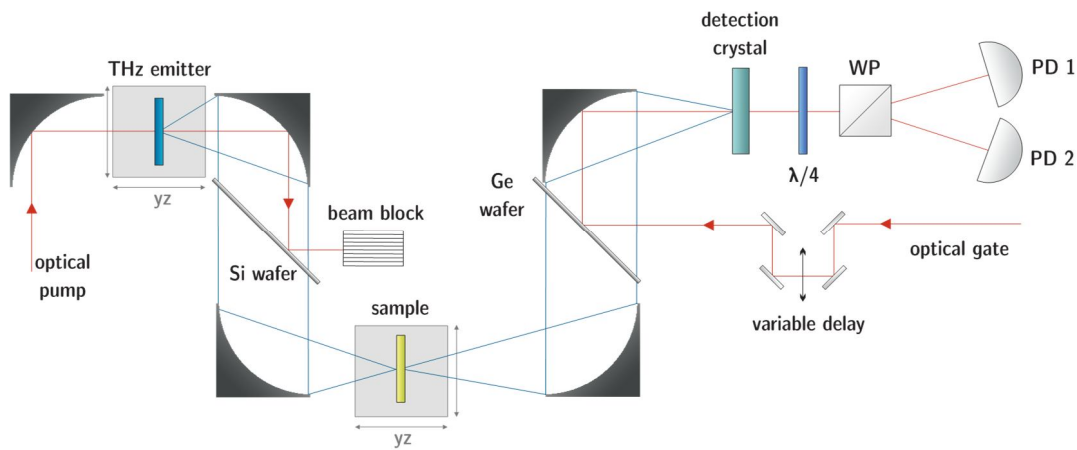

Figure 3. A typical THz time-domain spectrometer. An optical pump pulse is focused onto a THz emitter. The emitted THz wave is collimated by an off-axis parabolic mirror and subsequently focused onto the sample. A Ge wafer is used to block the residual pump pulse. After recollimation, the optical-gate beam is combined with the THz beam by using a Ge wafer. THz and gate beams

are focused into the electrooptic detection crystal. By delaying the two pulses with respect to one another, the THz-field-induced birefringence, a measure of the THz-electric-field strength in the detection crystal, is mapped out using a balanced detection scheme consisting of a quarter-wave plate (λ/4), a Wollaston prism (WP) and two photodiodes (PD1 and PD2).

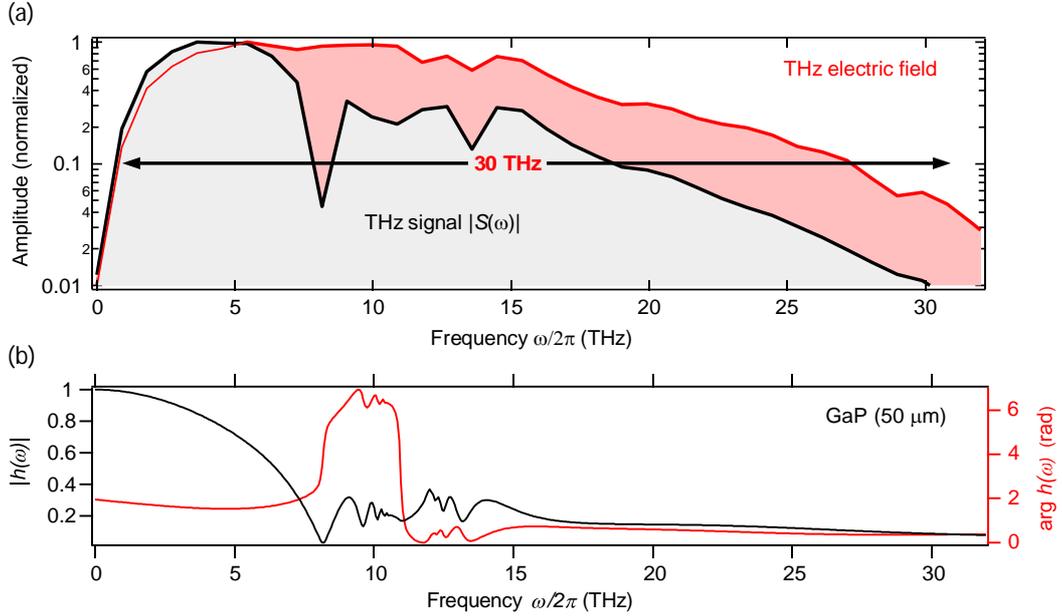

Figure 4. Typical data obtained from a spintronic THz emitter with a THz time-domain spectrometer (TDS). (a) Amplitude spectrum $|S(\omega)|$ (black curve) of the electrooptic signal of a THz pulse from a FM|NM stack and the corresponding extracted THz electric field spectrum $E_{\text{Det}}(\omega)$ (red curve) at the position of the 50-μm-thick GaP detector. (b) Complex-valued response function $H_{\text{EOS}}(\omega)$ of the 50-μm-thick GaP electrooptic crystal for the case of a 10 fs, 800 nm gate pulse used for THz detection [64].

Note that EO sampling (EOS) does not record the THz electric field directly. Instead, the measured THz signal $S(t)$ is determined by the convolution of the THz electric field $E_{\text{Det}}(t)$ right in front of the detector with the response function $H_{\text{EOS}}(t)$ of the THz detection process. In the frequency domain, $S(\omega)$ and $E_{\text{Det}}(\omega)$ are connected by a multiplication [64],

$$S(\omega) = H_{\text{EOS}}(\omega) E_{\text{Det}}(\omega) \,. \tag{4}$$

Fig. 4 shows the spectra $E_{\text{Det}}(\omega)$ and $S(\omega)$ of a typical THz pulse emitted from a STE upon pumping with ~10 fs optical laser pulses and detection using a 50-μm-thick GaP EO detection crystal. The minimum of $|S(\omega)|$ at ~8 THz is due to $H_{\text{EOS}}(\omega)$ (Fig. 4b) and results from the compensation of purely electronic and Raman-type contributions involving the crystal lattice to the EO effect in GaP [97]. As with the THz-generation process discussed above, by increasing the EO crystal thickness, the detected signal $S(t)$ increases in amplitude, but the bandwidth of $H_{\text{EOS}}(\omega)$ is reduced simultaneously because of the increased THz-field attenuation and phase mismatch between the THz and the probe pulse. Therefore, in the study of the STE, thin GaP is a suitable choice as the EO detector due to its relatively large bandwidth (see Fig. 4b) and sufficient signal strength.

2.3 *Typical THz-emission signals from STEs and their optimization*

Based on our model [Eqs. (3a), (3b) and (3c)], we now turn to implementations of STEs based on various material systems and mechanisms, which can strongly influence their performance and properties. We present the different optimization steps that, in hindsight, provided valuable insights into the underlying STE physics.

2.3.1   *Typical THz-emission signals from STEs*

In 2013, THz emission from metallic spintronic FM|NM heterostructures was reported [63]. The THz waveforms emitted from the studied Fe|Au and Fe|Ru thin films differed in polarity and dynamics, yet both THz signals reversed entirely upon reversing the in-plane sample magnetization (Fig.5a). The corresponding THz emission spectrum of Fe|Au peaked at 4 THz and had a bandwidth of about 10 THz (Fig. 5b), which was much broader than the spectrum obtained from Fe|Ru. This finding indicated different dynamics of the ultrafast currents in Au and Ru. *Ab-initio* calculations provided an explanation of the different dynamics within the superdiffusive spin-transport model (Section 3.1.1) [98]. This work identified a prototype FM|NM STE based on the ISHE and showed the potential of THz-emission spectroscopy for the study of ultrafast spin-related transport dynamics.

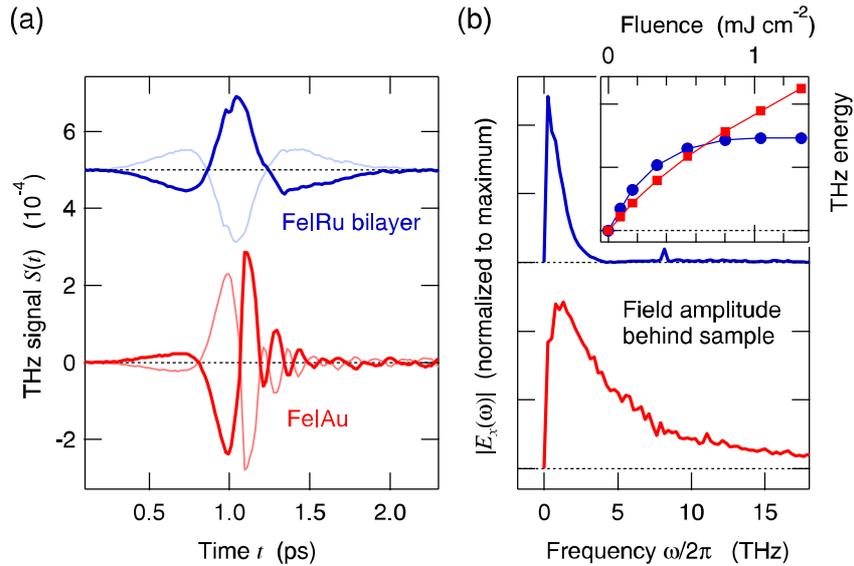

Figure 5. Experimental demonstration of THz emission in spintronic FM|NM stacks. (a) Emitted THz waveforms from Fe|Ru and Fe|Au thin-film heterostructures for opposite sample magnetizations. (b) The Fourier transform of (a), the inset is the pump-fluence-dependence of the

THz energy. Reproduced with permission from Nature Nanotechnology, 2013. 8: p. 256-260. Copyright 2016 Macmillan Publishers Limited.

Subsequent work [64] put Eq. (3a) to test by various experimental checks. First, the THz signal was found to reverse upon reversing the sample magnetization (Fig. 6a) [63, 64]. Second, by growing the FM|NM bilayer in reversed order, i.e., NM|FM, the polarity of the detected THz signal reversed because $j_s$ flowed from FM to NM and, thus, changed its sign (Fig.6b). Third, a sinusoidal dependence of the THz amplitude on the direction of the external magnetic field was found when a THz polarizer was inserted into the optical path, which demonstrated that the emitted THz pulse was linearly polarized with the THz electric field perpendicular to **M** (Fig. 6c). Fourth, the ISHE was confirmed as the dominant S2C mechanism by comparing the sign and the amplitude of the THz emission for different NM layers [64] (see Section 2.3.3).

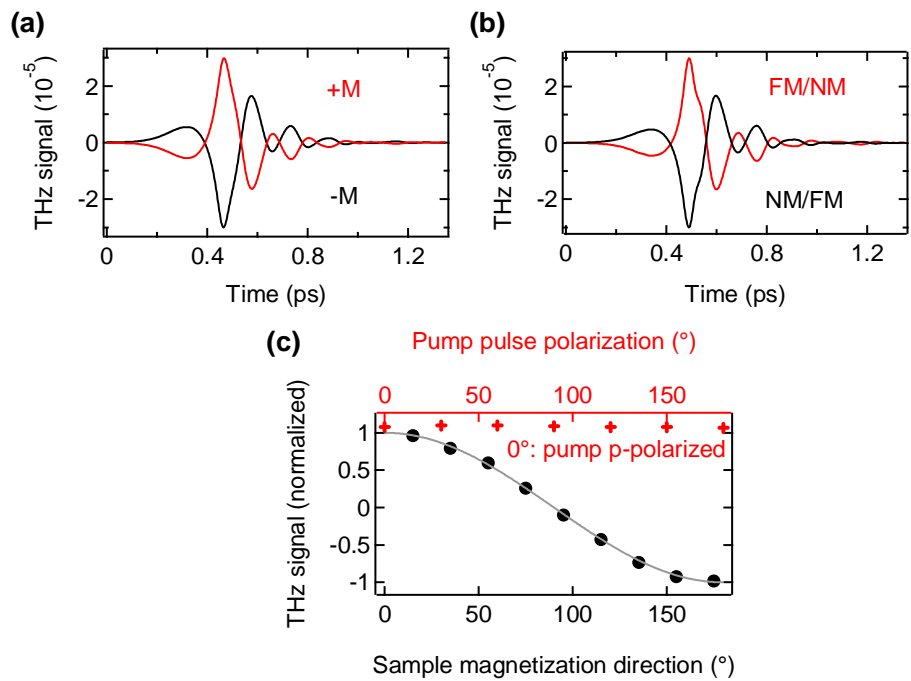

Figure 6. THz-signal symmetries from STEs consisting of FM|NM thin-film stacks with large spin-orbit coupling. (a) Polarity dependence of the detected THz signal on the direction of the sample magnetization **M**. (b) Detected THz-emission signals from FM|NM and NM|FM bilayers, i.e., with reversed growth order. (c) Dependence of the THz emission signal on the angle of the external magnetic field in the sample plane with a polarization-sensitive THz detection (black) and on the polarization direction of the linearly polarized pump pulse (red). Reproduced with permission from Nature Photonics, 2016. 10: p. 483. Copyright 2016 Macmillan Publishers Limited.

Following these checks, an in-depth optimization of the STE's geometry and materials was conducted, as detailed in the following subsections. The resulting optimized STE trilayer exhibits a remarkable performance as a THz source [64], as compiled in Section 1.3. In particular, the electric field (Fig. 7a) and bandwidth (Fig. 7b) of the emitted THz pulse from the STE right in front of the EO detector were compared to other commonly-used pulsed THz sources. The results highlight the high THz-emission efficiency from the STE, as well as its ultrabroad continuous spectrum. The extracted electric-field bandwidth $\Delta\omega_{THz}^{10\%}/2\pi$ at the detector position reaches about ~30 THz when using 10 fs pump pulses centered at a wavelength of 800 nm (Fig. 7). Therefore, spintronic multilayers are a novel ultrabroadband THz source with excellent performance and offer interesting perspectives for implementing future spintronic developments aiming at, for instance, even larger S2C performances [99].

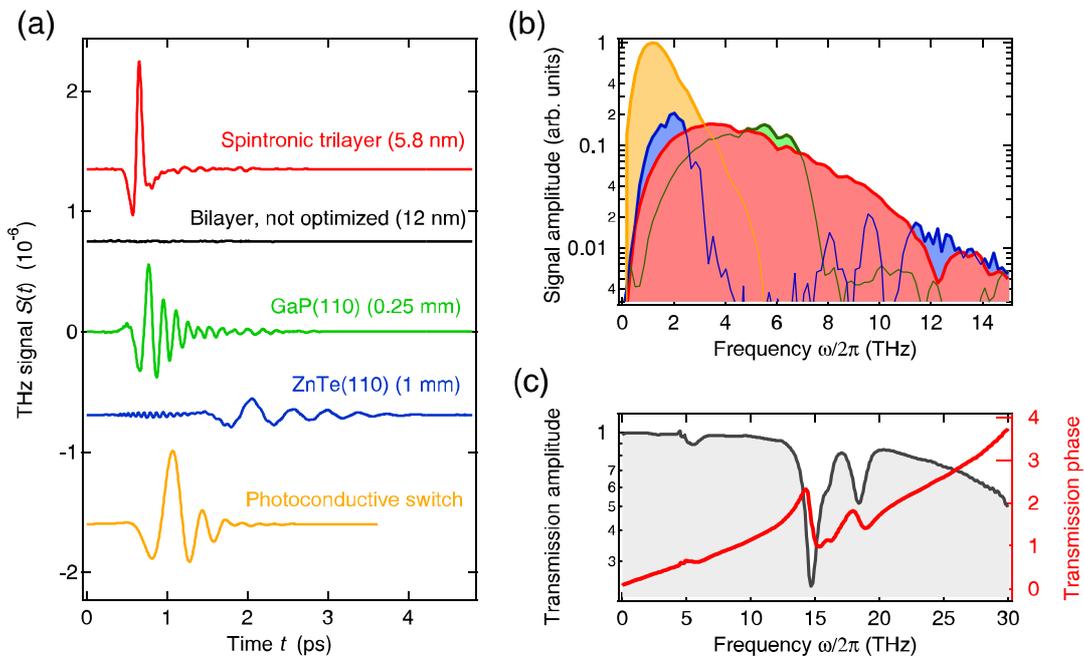

Figure 7. Comparison of the emitted THz waveforms from different THz emitters under identical conditions in (a) the time domain and (b) the frequency domain employing a poled-polymer electrooptic detector. (c) Spectral amplitude and phase of the THz transmission of a 7.5-μm-thick PTFE thread-seal tape measured with a spintronic emitter using a 10-μm-thick ZnTe electrooptic sensor. Reproduced with permission from Nature Photonics, 2016. 10: p. 483. Copyright 2016 Macmillan Publishers Limited.

### 2.3.2 *Optimization of the STE THz-emission amplitude*

The STE has the advantages of simple device structure, easy fabrication, adjustable polarization and ultrabroad bandwidth, making it a promising THz source. However, to widen its applications, its THz-generation efficiency needs to be optimized. In recent years, several studies aimed at optimizing the STE in terms of THz-emission efficiency [64-67, 69, 71, 73, 94, 95, 100-106], the ease-of-use and compatibility [68, 107-109], as well tuneability of the THz-emission properties such as the polarization [72, 74, 110, 111]. From such optimization studies, one might not only expect an enhanced STE performance but also a more in-depth understanding of the physical phenomena governing the STE operation.

To analyze the STE performance, we start with Eqs. (3a), (3b) and (3c) and make the following assumptions: The spatial profile of the spin current $j_s$ is calculated from spin-diffusion equations [112], the spin-current amplitude is assumed to scale with the energy density deposited by the pump pulse, and S2C is dominated by the ISHE in the NM layer. Thus, for metal films that are thin compared to the THz wavelength and the THz as well as the optical penetration depth, i.e., for film thicknesses well below 20 nm, the emitted THz electric field can be described in the plane-wave approximation by [64, 105]

$$E_{\text{THz}} = \frac{AF_{\text{inc}}}{d_{\text{NM}} + d_{\text{FM}}} \cdot j_s^0 \, t_{\text{FM/NM}} \lambda_{\text{NM}} \tanh \frac{d_{\text{NM}}}{2\lambda_{\text{NM}}} \cdot \gamma \cdot \frac{eZ_0}{n_1 + n_2 + Z_0 G}. \quad (6)$$

$$\quad\quad\quad (1) \quad\quad\quad\quad (2) \quad\quad\quad\quad (3) \quad\quad\quad (4)$$

Here, the terms (1)-(4) correspond to the different model steps introduced in Sect. 2.1: (1) pump-pulse absorption, (2) spin-current generation, (3) spin-to-charge current conversion and (4) charge-current-to-electric-field conversion. In Eq. (6), the quantity $A$ is the absorbed fraction of the incident pump-pulse fluence $F_{\text{inc}}$, $j_s^0$ is the generated spin-current density per pump-pulse excitation density, $t_{\text{FM/NM}}$ is the interfacial spin-current transmission amplitude between FM and NM layers, $\lambda_{\text{NM}}$ is the spin-current relaxation length in the NM layer, and $\gamma$ is the NM spin Hall angle. The quantities $n_1$, $n_2$, $e$ and $G$ were already introduced following Eqs. (3b) and (3c). All parameters except $d_{\text{NM}}$, $e$, $Z_0$ and those in term (1) depend, in principle, on the THz frequency. Table 1 summarizes typical values of the parameters that enter Eq. (6) found in optimized STEs (see Sect. 2.3.2).

A recently introduced model [113] that involves spin voltages (see Section 3.1.3) allows one to derive a relation between material-specific parameters and $j_s$, which reads as

$$j_s(t) \propto \int d\tau \, I(\tau) H_{j_s I}(t - \tau) \text{ with } H_{j_s I}(t) = \Theta(t)\big[A_{\text{es}} e^{-\Gamma_{\text{es}} t} - A_{\text{ep}} e^{-\Gamma_{\text{ep}} t}\big]. \quad (7)$$

Here, $I(t)$ is the pump-pulse intensity envelope, $\Theta$ is the Heaviside step function, $\Gamma_{\text{es}}^{-1}$ and $\Gamma_{\text{ep}}^{-1}$ are the time constants of electron-spin and electron-phonon equilibration, respectively, $A_{\text{es}} = (\Gamma_{\text{es}} - R\Gamma_{\text{ep}})/(\Gamma_{\text{es}} - \Gamma_{\text{ep}})$ and $A_{\text{ep}} = (1 - R)\Gamma_{\text{ep}}/(\Gamma_{\text{es}} - \Gamma_{\text{ep}})$, and $R$ is the ratio of the electronic and total heat capacity of the sample. In essence, the spin-current dynamics inside the STE [114] are determined by a rise time given by the pump-pulse duration and a decay that has contributions of $\Gamma_{es}^{-1}$ (typically 100 fs) and $\Gamma_{ep}^{-1}$ (typically several 100 fs) [113]. This result implies that the bandwidth of the STE is ultimately only limited by the pump-pulse duration and by the time it takes to absorb a pump photon.

In summary, the THz-emission efficiency depends strongly on the material properties of the FM and NM layers, as well as the photonic and geometric design of the STE. The impact of all these parameters on the STE performance will be discussed in the following.

| Parameter | $A$ | $F_{inc}$ | $d_{NM}$ | $d_{FM}$ | $t_{FM/NM} j_s^0$ | $\lambda_{NM}$ | $\gamma$ | $Z$ |
|---|---|---|---|---|---|---|---|---|
| Typical value | 0.5 | 0.1 $\frac{mJ}{cm^2}$ | 3 nm | 3 nm | $10^{32} \frac{1}{s\,m^2}$ | 2 nm | 0.1 | 50 Ω |

Table 1. Typical frequency-averaged parameter values of optimized STEs (see Eq. 6) [64, 115-117]. To calculate the value of $Z$, typical values of $n_1 = 1$, $n_1 = 2.5$ and $G = 4$ S were assumed.

### 2.3.3 *Material choice*

In view of Eq. (6), intrinsic sample parameters that impact the THz generation efficiency are $\gamma$, $\lambda_{rel}$, $t_{FM/NM}$, $j_s^0$, $A$ and $Z = Z_0/(n_1 + n_2 + Z_0 G)$.

*NM variation.* Seifert et al. [64] performed systematic studies of the STE's NMs including Cr, Pd, Ta, W, Ir, Pt$_{38}$Mn$_{62}$ and Pt. It was found that the emitted THz amplitude was highly dependent on the NM (Fig. 8). Compared to FM|Pt, the THz-emission signals with opposite sign from FM|Ta and FM|W originated from their negative spin Hall angle $\gamma$ as confirmed by *ab-initio* calculations. Related works [118] could confirm these results [64] and extended the NMs to Au [119], Ru [63, 102], Al [120], IrMn$_3$ [121] and Mn$_2$Au [122].

An interesting approach to enhance the S2C amplitude is by alloying, which might enhance the skew-scattering contribution to the S2C significantly. Accordingly, Gueckstock and coworkers performed a systematic study that showed the profound impact that alloying layers at the FM/NM interface can have on the STE performance [115]. Up to now, however, Pt-based STEs that rely on the ISHE yield the highest THz-emission amplitudes.

For the IREE-based STE, the range of the studied material systems is rather limited. Up to date, only Ag/Bi interfaces [117, 123], monolayer MoS$_2$ [124] and the surface of Bi$_2$Se$_3$ [125] were investigated. Fe|Ag|Bi and Co|Bi$_2$Se$_3$ stacks showed a THz-emission

efficiency of about 1/5 of ZnTe [117, 123, 125]. Previous DC-transport studies indicated a large potential in terms of S2C for IREE systems [126]. Exploiting this promising feature for STEs implies maximizing the Rashba S2C parameter $\lambda_{\text{IREE}}$ (see Sect. 3.2.2), which has the unit of length and is the equivalent to the term $\lambda_{\text{NM}} \gamma \tanh(d_{\text{NM}}/2\lambda_{\text{NM}})$ for the ISHE-based STE [see Eq. (6)]. However, experiments could not yet demonstrate the expected large THz-emission efficiencies for IREE-based STEs, which might indicate limitations in terms of $t_{\text{FM/NM}}$. In another approach, THz emission from ISHE and IREE could be implemented in the same device. Adding the two respective THz signals could boost the efficiency to a much higher level in future STE designs [117].

*Metallic FM variation.* The choice of the FM material for the STE was also investigated [64], and the results indicated that Co, Fe or the binary alloys containing Ni, Fe or Co as the FM layer have a much higher THz-emission efficiency than pure Ni (Fig. 9a). The lower THz-emission efficiency of Ni-based STEs might be related to its lower Curie temperature or a lower value of $t_{\text{FM/NM}}$ [90]. Related works studied the detailed composition of Co and Fe in $(Co_xFe_{1−x})_{80}B_{20}$|Pt structures and found that the THz-emission efficiency is maximized at $x = 0.1\text{-}0.3$ [66, 104] or in the range of $x = 0.5\text{-}0.8$ for $Co_xFe_{1−x}$|Pt [127]. A temperature-dependent study found a minor impact on the THz-emission amplitude in Co|Pt structures between 10 and 300 K [128], consistent with a related study in Co|Mn$_2$Au [122]. Among the studied samples, sputter-deposited $Co_{40}Fe_{40}B_{20}$ is the most efficient FM for the STE so far.

In the future, half-metallic FM systems such as Heusler alloys [129-132], which promise large spin polarizations, might help increase the STE performance. However, it should be noted that it is not the value of the spin polarization alone that determines the STE performance, but that also the change of magnetization with electronic temperature has a major impact [113].

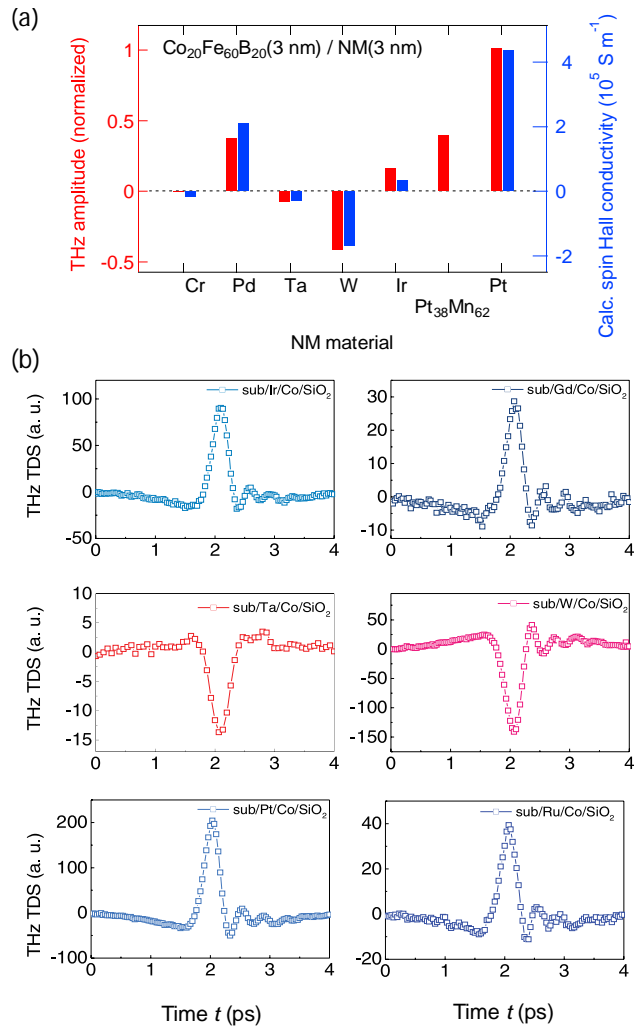

Figure 8. Comparison of THz-emission amplitudes from FM|NM heterostructures based on the inverse spin Hall effect for different NMs. (a) THz emission from CoFeB|NM with different NMs (red, normalized to CoFeB|Pt) along with *ab-initio* calculations of the spin Hall conductivity. Reproduced with permission from Nature Photonics, 2016. 10: p. 483. Copyright 2016 Macmillan Publishers Limited. (b) THz emission from Co|NM with different NMs. The figures are adapted from Refs. [64, 66]. Reproduced with permission from Adv Mater, 2017. 29: p. 1603031. Copyright 2017 Wiley-VCH.

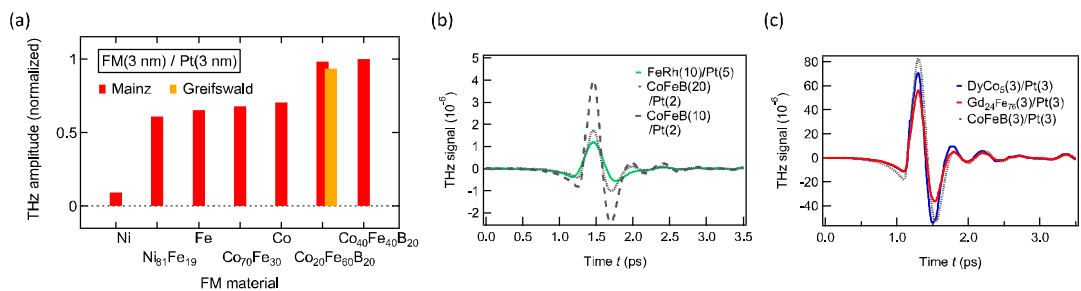

Figure 9. Comparison of THz-emission amplitudes from FM|NM heterostructures based on the inverse spin Hall effect for different FMs (a) THz emission from FM|Pt with different FM materials grown in different labs. Reproduced with permission from Nature Photonics, 2016. 10: p. 483. Copyright 2016 Macmillan Publishers Limited. (b)-(c) THz emission from

heterostructures containing complex magnets such as $DyCo_5$, $Gd_{24}Fe_{76}$ or FeRh and Pt as the ISHE material. The figures are adapted from Refs. [64, 133]. Seifert, T., U. Martens, S. Gunther, et al., Spin., 7, p. 1740010, 2017; licensed under a Creative Commons Attribution (CC BY) license.

*Beyond ferromagnets.* As an interesting alternative to FMs, ferrimagnets (FIM) [73, 133-138], antiferromagnets [73, 133, 139-141], and magnetic insulators (MI) [88, 142] were studied as the spin-current-generating magnetic layer in the STE. Seifert et al. compared the THz-emission signals from CoFeB|Pt and M|Pt samples, where M = $DyCo_5$, $Gd_{24}Fe_7$, $Fe_3O_4$ and FeRh (Fig 9b, 9c and Ref. [133]). A parameter, defined as the figure of merit $\text{FOM}_M = \left\| t_{\text{FM/NM}} j_s^0 \right\|_M / \left\| t_{\text{FM/NM}} j_s^0 \right\|_{\text{ref}}$, was introduced to compare the spin current generated in a M|Pt bilayer with that of a CoFeB|Pt reference sample, for which $\text{FOM}_{\text{CoFeB}} = 1$ by definition. For the metallic FIM $DyCo_5$ and $Gd_{24}Fe_7$, the corresponding FOMs were extracted to be 0.85 and 0.79, respectively. These observations were ascribed to the reduced saturation magnetization due to the ferrimagnetic order. The spin polarization of $DyCo_5$, $Gd_{24}Fe_7$ and CoFeB is 40%, 36% and 65%, respectively. Evidently, the FM order plays an important role in the performance of the STE.

Moreover, the experimental data of $Fe_3O_4$|Pt gave FOM = 0.09, which was attributed to a low value of $t_{\text{FM/NM}} j_s^0$. These results suggest that low-conductivity magnetic materials are not suitable for efficient STEs, and that it is essential for the spin-current-generating layer to have a metal-like conductivity. Furthermore, the THz emission from FeRh|Pt, was found to be strongly temperature-dependent, which can be attributed to the antiferromagnet-to-FM phase transition of FeRh around room temperature [133, 143]. Although these alternative structures have not yet outperformed the FM-based STE, interesting new functionalities might arise. For instance, it was shown that the spin current is governed by only one the magnetic sublattices in an alloyed ferrimagnet such as GdFeCo [134].

*Magnetic-insulator/NM heterostructures.* The spin-current generation and injection in MI|NM bilayers is based on a different mechanism compared to the fully metallic STE. A typical example is a ferrimagnetic-insulator|NM (FII|NM) stack. Because of the insulating nature of the magnetic layer, no conduction electrons are available to carry a spin current from the FII to the NM. Instead, the spin transport is mediated by magnons inside the FII.

In the common DC version of the so-called spin Seebeck effect (SSE), a temperature gradient throughout the FII|NM sample leads to a magnon accumulation at the FII/NM interface. In contrast, the ultrafast SSE relies on the selective heating of the NM layer by the femtosecond laser pulse creating a step-like temperature profile across the FII/NM

interface. The ultrafast increase of electronic temperature in the NM layer leads to an enhanced scattering of NM electrons off the interface. During these scattering events, the NM electrons couple to the magnetic order in the FII through the interfacial exchange interaction. Through a second-order effect, the pump-induced enhanced exchange-scattering noise inside the NM gets rectified, giving rise to a net $\mathbf{j}_s$. The spin current is polarized along the FII magnetization and flowing perpendicular to the interface.

The ultrafast SSE was experimentally observed in ferrimagnetic YIG|Pt bilayers by means of optical pump-probe spectroscopy with ~1 ps time resolution [144] and THz emission spectroscopy with 27 fs resolution [88], confirmed for different YIG thicknesses [114] and extended to low temperatures [145]. The observed THz emission signal was of magnetic origin and related to a spin current inside the NM followed by the ISHE [see Fig. 10 and Eq. (3a)]. Notably, the ultrafast SSE can coexist with the spin-voltage-driven spin transport in magnetic materials such as the half-metallic ferrimagnet $Fe_3O_4$ (magnetite) [114]. In the future, the ultrafast SSE may provide an all-optical access to study magnons in FIIs and enable the generation of ultrafast spin currents using MIs [146]. In terms of THz-emission strength, the SSE scenario in FII|NM samples is about 2 to 3 orders of magnitude less efficient than the spin-voltage mechanism in all-metallic structures.

Another progress in this field is the THz generation in antiferromagnet insulator|NM (AFMI|NM) stacks, such as NiO|Pt bilayers [139]. In the case of NiO, the spin-current generation was ascribed to the inverse Faraday effect [48]. Although its efficiency is much lower than for the fully metallic STEs, this approach demonstrates the potential of AFMIs and AFMs, in general, in THz spintronics [147, 148].

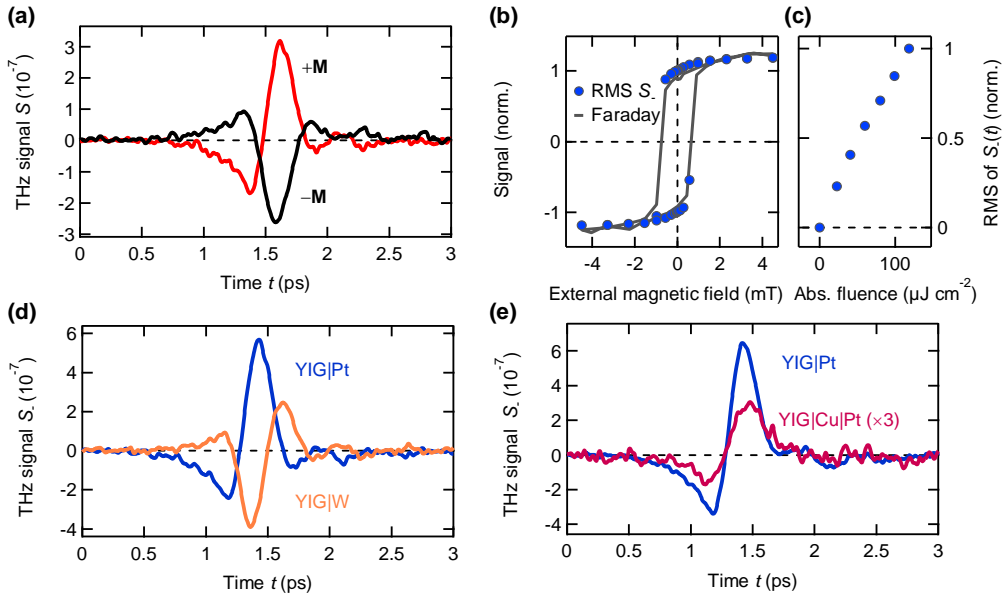

Figure 10. THz emission of photoexcited YIG|Pt heterostructures. (a) THz emission signals $S(t, \pm\mathbf{M})$ from a YIG|Pt sample for opposite directions of the in-plane YIG magnetization $\mathbf{M}$ as

a function of time $t$. (b) Amplitude of the THz signal $S_-(t) = S(t, \mathbf{M}) - S(t, -\mathbf{M})$ and the Faraday rotation of a continuous-wave laser beam (wavelength 532 nm) as a function of the external magnetic field. (c) Fluence dependence of the THz-emission amplitude. (d) THz-emission signal from YIG samples capped with Pt and W. (e) THz emission signal from a YIG film capped with Pt or Cu|Pt. The figure is adapted from Ref. [88]. Seifert, T.S., Jaiswal, S., Barker, J. et al., Nat Comm., 9, 2899, 2018; licensed under a Creative Commons Attribution (CC BY) license.

### 2.3.4 *Thickness optimization*

The thickness of each layer in the STE's FM|NM heterostructure can strongly affect the THz-emission efficiency. According to Eq. (6), $d_{\mathrm{NM}}$ and $d_{\mathrm{FM}}$ should be optimized with respect to the factor $A$, the total metal-film conductance $G = \int_0^{d_{\mathrm{NM}}+d_{\mathrm{FM}}} \mathrm{d}z\sigma(z)$, and the spin-current-propagation factor $\lambda_{\mathrm{NM}} \tanh(d_{\mathrm{NM}}/2\lambda_{\mathrm{NM}})$.

Notably, the factor $A$ was found to be approximately independent $d_{\mathrm{NM}}$ and $d_{\mathrm{FM}}$ for total thicknesses in the range from 3-10 nm [64]. For larger thicknesses, however, $A$ follows an exponential decay plus an offset [64, 105].

In terms of the spin-current propagation, the factor $\tanh(d_{\mathrm{NM}}/2\lambda_{\mathrm{NM}})$ in Eq. (6) can minimize the THz-emission amplitude significantly if $d_{\mathrm{NM}} < \lambda_{\mathrm{NM}}$, because multiple spin-current reflections superimpose destructively [105]. If, on the other hand, $d_{\mathrm{NM}} \gg \lambda_{\mathrm{NM}}$, the spin current has decayed before reaching the NM/substrate or NM/air boundary. Thus, an optimal total thickness for the STE exists, which maximizes the THz-emission amplitude. One can estimate that this optimum occurs at $d_{\mathrm{NM}} \sim 2\lambda_{\mathrm{NM}}$, which corresponds to the peak in Fig. 11a. Therefore, the optimized thickness of the NM layer should be about 2 nm for Pt ($\lambda_{\mathrm{NM}} = 1$ nm) and in a similar range for other NMs with large S2C efficiency [64, 105, 106].

Importantly, for DC spin currents, $\lambda_{\mathrm{NM}}$ in Eq. (6) equals the spin-diffusion length, but is expected to be shortened and approach the electron mean free-path length $\lambda_{\mathrm{mfp}}$ for THz spin transport [92]. This prediction was confirmed recently by measuring spin-current propagation lengths at GHz and THz rates, revealing a 4-fold reduction in the case of THz spin currents [149]. Accordingly, the measured spin-diffusion length of Pt is of the order of 10 nm [150], whereas, for THz spin currents, a $\lambda_{\mathrm{NM}}$ of approximately 1 nm was found for thin Pt films. Similar values were reported for other ISHE materials [91, 105]. This observation allows one to draw the following important conclusion: For the typical metallic thin films used for spintronic THz emission, the THz conductivity is well described by the Drude model [151], and one often finds to a good approximation $\sigma_{\mathrm{NM}}(\omega) \approx \sigma_{\mathrm{NM}}(\omega = 0)$ due to the small Drude scattering time $\tau$ in the range of a few femtoseconds [79, 105]. This fact directly implies that $\sigma_{\mathrm{NM}} \propto \tau \propto \lambda_{\mathrm{mfp}} = \lambda_{\mathrm{NM}}$.

We can go one step further and address the impact of the above argumentation on the emitted THz electric field $E$. For typical NM film thicknesses of a few nanometers, one typically has $\tanh(d_{NM}/2\lambda_{NM}) \approx 1$ and $n_1 + n_2 + Z_0 G \approx n_1 + n_2$. Eq. (6), thus, implies that $E \propto \sigma_{NM}\gamma$. In other words, $E$ is rather proportional to the spin Hall conductivity $\sigma_{NM}^{SH} = \sigma_{NM}\gamma$ and not to the spin Hall angle $\gamma$ [64, 91]. Note that this simple relationship only holds in cubic/isotropic materials such as the typical solids used for STEs. It might be more complex otherwise [152].

Similar to $d_{NM}$, the thickness of the FM layer has a profound impact on the THz-emission amplitude, too (Fig. 11b) [117]. Here, a critical minimal thickness $d_0$ of typically about 0.5 nm exists, below which the magnetic order of the FM layer is not well established [153], as also confirmed by recent magneto-optical Kerr effect studies [65]. The impact of this magnetically dead layer can be described by a correction factor $\tanh[(d_{NM} - d_0)/\lambda_{FM}]$ to Eq. (6) [95], where $\lambda_{FM}$ accounts for the finite spin propagation length inside the FM layer, i.e., the region of the FM that still contributes to the spin current into the NM as shown in Fig. 2 [95]. Based on recent studies [64-66, 106, 154], the optimized thickness of the FM layer (FM = Ni, Co, Fe, CoFeB) is 1-4 nm.

To summarize, thickness-dependent studies of the FM and the NM layers show that the THz spin and charge currents are localized within a few nanometers at the FM/NM interface and can, thus, be considered interfacial probes of ultrafast charge and spin dynamics.

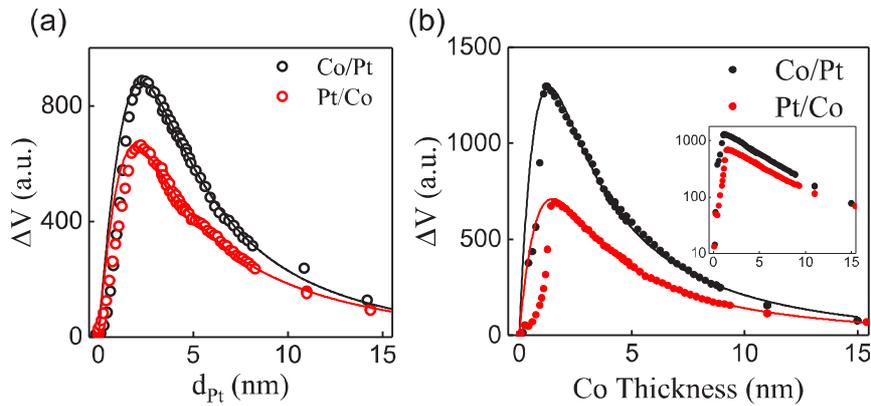

Figure 11. Dependence of THz-emission amplitude on the thickness of (a) the NM and (b) the FM layers under identical experimental conditions. Reproduced with permission from Phys. Rev. Lett., 121, p. 086801, 2018. Copyright 2018 American Physical Society.

### 2.3.5 *Crystal structure, degree of disorder and alloying*

Controlling the atomic arrangement and degree of disorder in the bulk and at the interfaces of the STE's metallic layers might optimize the generation ($j_s^0$), the injection efficiency ($t_{FM/NM}$) and propagation ($\lambda_{NM}$) of the spin current as well as the two parameters $A$ and $Z$.

Li and coworkers systematically studied the influence of interfacial roughness, crystal structure and interface intermixing on the THz emission from polycrystalline Co|Pt heterostructures [100]. As shown in Fig. 12a, the THz emission based on the spin voltage (see Sect. 2.1 and 3.1) strongly depended on the interfacial roughness and could be largely improved by increasing the smoothness of the samples. A possible reason is that the spin-flip probability is strongly correlated with the roughness and/or other types of disorder at the Co/Pt interface. It is argued that more disorder decreases the spin-current injection across the Co/Pt interface, i.e., the factor $t_{FM/NM}$ in Eq. (6).

However, the same study reported that the THz-emission based on the inverse spin-orbit torque (ISOT, Fig. 12b, see Sect 2.3.10) increases first within the same sample for small degrees of disorder before also decreasing for larger disorder. This surprising behavior might be attributed to the interfacial origin of the ISOT mechanism. In contrast, an intermixing layer at the interface, i.e., a $Co_xPt_{1-x}$ spacer layer, resulted in an increased ISHE-based THz-emission amplitude. At least two explanations are possible: (i) The $Co_xPt_{1-x}$ alloy layer increases $t$, or (ii) an additional IREE arises from the $Co_xPt_{1-x}$ alloy layer. Clearly, these results indicate that a suitable interlayer can control the THz-emission efficiency. Similar conclusions were reached in a related work [103] that reported a strong impact of electron-defect scattering on the STE performance by comparing Fe|Pt bilayers with different degrees of crystallinity. Along these lines, epitaxial Fe films were shown to exhibit an anisotropic THz emission performance, which was attributed to strain-induced effects [155].

Another work revealed the strong influence that interfacial intermixing can have on the STE performance [115]. In this study, Gueckstock et al. showed that the order, in which the layers are deposited during the sputtering growth process, determines the sign of the S2C in the interfacial alloy layer (see Sect. 3.2).

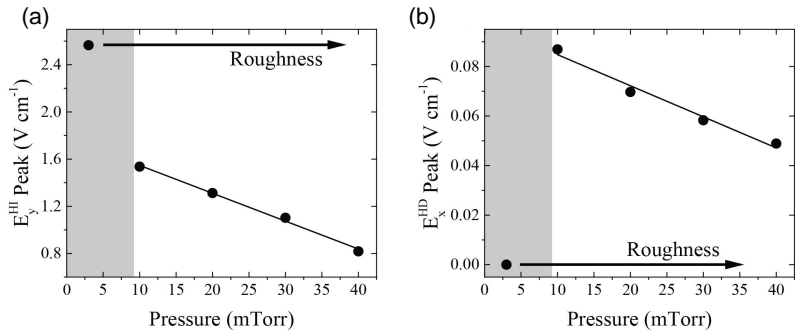

Figure 12. Influence of interfacial roughness on (a) the spin-voltage-based and (b) the THz-emission amplitude related to the inverse spin-orbit torque from Co|Pt. Reproduced with permission from Phys. Rev. Mat., 3, p. 084415, 2019. Copyright 2019 American Physical Society.

The crystal structure of FM|NM layers can be further optimized by post-annealing of sputter-grown STEs [94, 104]. As seen in Fig. 13, in CoFeB|Ta heterostructures [94], the annealing process had different effects: (a) B atoms diffused into the Ta layer, which enlarged the S2C efficiency and decreased $\lambda_{NM}$. (b) An enhanced crystallization of the CoFeB layer can have a profound impact on the electronic transport properties. (c) The saturation magnetization decreased because of the atomic mixing between CoFeB and Ta, which diminished $j_s^0$ and likely also impacted $t_{FM/NM}$. The best THz-emission performance for CoFeB|Ta bilayers was found for a post-growth annealing at 300 °C for 1 h. On the other hand, THz emission from $Co_xFe_{1-x}B_{20}$|Ta bilayers [104] was reported to peak for $x = 0.2$ and an annealing temperature of ~400 °C, where the largest saturation magnetization of CoFeB was obtained.

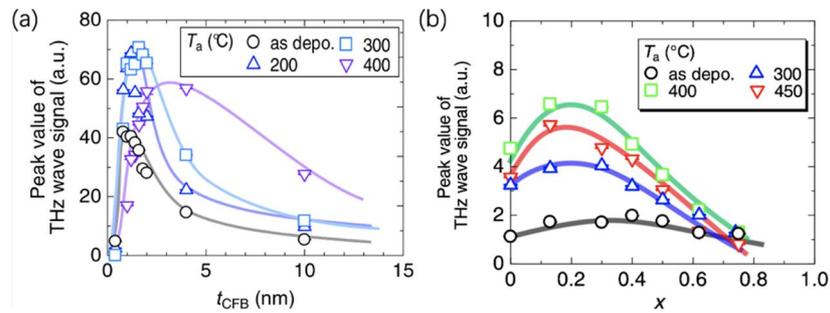

Figure 13. Effect of post-growth annealing on the THz emission from spintronic THz emitters based on (a) CoFeB|Ta for different CoFeB thickness $t_{CFB}$ and (b) $(Co_xFe_{1-x})_{80}B_{20}$|Ta for varying Co-content $x$. Reproduced with permission from Phys. Rev. B 100, 140406(R), 2019. Copyright 2019 American Physical Society. Reproduced from Appl. Phys. Lett. 111, 102401, 2017, with the permission of AIP Publishing

### 2.3.6 *Impact of the substrate*

According to Eq. (6), the substrate, on which the STE is grown, impacts the THz emission directly through the impedance $Z$. Thus, spectral features in the THz emission can arise, which are footprints of the substrate phonon modes [75]. Moreover, the substrate material influences the heat transport away from the excited STE area, which becomes an important factor for STE excitation conditions with large pump powers [69, 156].

### 2.3.7 *Pump-pulse characteristics*

Specific pump-pulse parameters such as the wavelength, the duration and the temporal structure play a critical role for the STE performance. A major experimental challenge for wavelength-dependent studies is to maintain a constant pump-pulse duration for different center wavelengths. Following this approach, for 1550, 800 and 400 nm center wavelength [68, 157] and for the range 900-1500 nm [67], it was found that the structure of the emitted THz pulse is independent of the pump wavelength The majority of the recent studies [67, 68] also showed that the THz-emission efficiency is largely independent of the pump wavelength. One exception from this behavior [109] was ascribed to a wavelength-dependent absorptance of the STE.

From a fundamental-science viewpoint, the pump-wavelength independence of STEs is fully consistent with the spin-transport model of Eq. (7). From an applied viewpoint, it is distinctly different from conventional THz sources, e.g., nonlinear optical crystals and semiconductor-based THz emitters. It enables the integration of the STE into systems driven by low-cost and compact femtosecond fiber lasers without loss of efficiency.

Regarding the pump-pulse duration $\Delta t_{\text{pump}}^{50\%}$ (defined as the full-width at 50% maximum of the intensity envelope), experimental results [64, 69, 158] suggest that the product of THz bandwidth $\Delta \omega_{\text{THz}}^{10\%}/2\pi$ of the focused THz beam at the detector position (see Figs. 3 and 4) and $\Delta t_{\text{pump}}^{50\%}$ is a constant that is given by

$$\frac{\Delta t_{\text{pump}}^{50\%} \Delta \omega_{\text{THz}}^{10\%}}{2\pi} \approx 0.5. \qquad (8)$$

Here, $\Delta \omega_{\text{THz}}^{10\%}$ is defined as the full width at 10% of the maximal THz electric field amplitude in the spectral domain. The temporal structure of the pump pulse is a further degree of freedom that allows one to generate double or multiple THz pulses by exciting the STE with multiple time-delayed pump pulses [111, 159].

2.3.8 *Impact of nano- and microstructuring*

Aside from FM|NM bilayer structures, more complex STE designs were proposed to increase the THz-emission efficiency. The basic idea is to gain more THz electric field per pump-pulse energy by A) increasing $j_s$ and the S2C strength and by B) better photonic management, which includes an optimized pump-pulse absorption (absorptance $A$) and a maximized current-to-electric-field conversion efficiency (THz impedance $Z$).

Regarding A), a promising approach utilizes trilayer STEs with the structure NM$_1$|FM|NM$_2$ (Fig. 14a) [64, 69, 101]. This design takes advantage of the spin currents in the forward as well as the backward propagation direction. Accordingly, the spin currents injected into NM$_1$ ($j_s^1$) and NM$_2$ ($j_s^2$) have opposite directions. To avoid a cancelation of the net charge current (sum of $j_c^1$ and $j_c^2$ inside the two NM layers), the spin Hall angles of NM$_1$ and NM$_2$ should have opposite signs [see Eq. (3a)]. Indeed, measurements of the THz-emission amplitude from W|Co$_{40}$Fe$_{40}$B$_{20}$|Pt trilayers revealed a doubling as compared to the Co$_{40}$Fe$_{40}$B$_{20}$|Pt bilayer counterpart with identical total thickness [64].

Regarding B), to increase $A$ at the expense of $Z$, a multilayer structure [Pt|Fe|MgO]$_n$ was fabricated (Fig. 14b) [65], where $n$ is the index number of the repeating periods. A 2-nm-thick MgO layer is used to isolate the spin current from the neighboring Fe layers to avoid spin-current leakage into the adjacent bilayer unit. The maximum THz-emission amplitude was found for $n = 3$ with a ~70% increase of THz-emission amplitude in comparison to $n = 1$ for the bilayer STE. For $n > 3$, the lowered excitation density and the increased STE impedance can decrease the THz-emission efficiency [160].

The two approaches of STE stacking and trilayer design were combined for trilayer STEs based on W|Fe|Pt films. In Ref. [101], the THz emission from [W|Fe/Pt/SiO$_2$]$_n$ periodic structures for $n = 1, 2$ and 3 with different SiO$_2$ thickness ($d_{SiO}$) was measured. Controlling $d_{SiO}$ allows for the increase of the incident light absorption via minimizing the reflection at interfaces/surfaces, and hence boost the THz-emission efficiency. An optimized structure with $n = 2$ and $d = 100$ nm could be identified with a corresponding efficiency that is ~1.7 times larger than that for $n = 1$.

Note, however, that the best stacking index $n$ depends on the individual NM and FM layer properties and, thus, not for each trilayer STE, $n > 1$ results in an increased THz-emission amplitude (unpublished results by some of the authors).

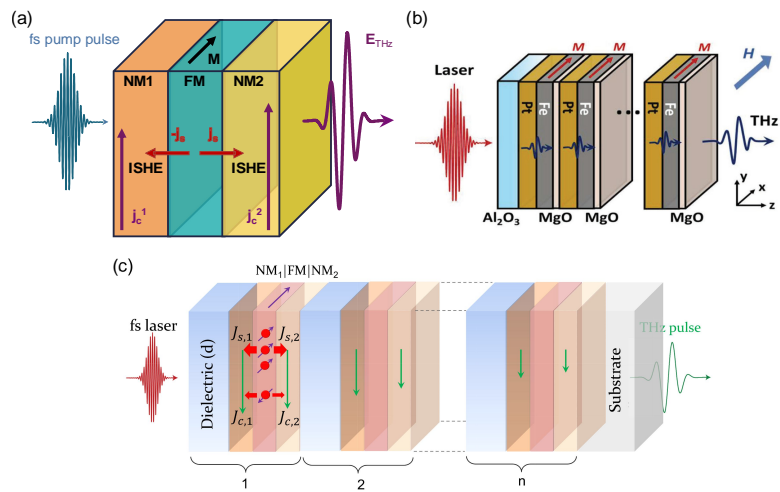

Figure 14. Structure design concepts for the spintronic THz emitter. (a) NM$_1$|FM|NM$_2$ structure, where NM$_1$ and NM$_2$ have spin Hall angles of opposite signs. Reproduced from Appl. Phys. Lett. 110, 252402, 2017, with the permission of AIP Publishing. (b) [Pt|Fe|MgO]$_n$ periodic structure. Reproduced with permission from Adv. Opt. Mat., 4, p.: 1944-1949, 2016. Copyright 2016 Wiley-VCH. (c) [W|Fe|Pt|SiO$_2$]$_n$ periodic structure. Reproduced with permission from Adv. Opt. Mat., 6, 1800965, 2018. Copyright 2018 Wiley-VCH.

The study of Ref. [111] arranged multiple STEs in one setup and superimposed the two emitted THz pulses. By controlling the external magnetic fields for the two STEs separately, the polarization of the combined THz pulse could be tuned. Clearly, the two THz emitters placed in a cascaded geometry can optimize the THz-generation efficiency by recycling the otherwise wasted reflected or transmitted fraction of the pump pulse. Similarly, future approaches might aim at combining the THz pulses that are propagating in forward and in backward direction with respect to the pump pulse propagation direction, which would easily increase the THz-emission amplitude by a factor of 2.

Spin-valve or synthetic antiferromagnet designs, that is, samples that contain multiple, possibly magnetically coupled FM layers with different magnetic properties such as coercive field strength offer another promising approach to enrich the STE functionality [161-163]. Alternatively, also magnetic tunnel junctions can be used to tune the spintronic THz emission [164].

Additionally, some works reported THz emission from microstructured FM|NM samples (also see Sect. 5.3), which allow one to control the emitted THz pulse shape and polarization to some extent [65, 71, 72, 165-168].

2.3.9 *Upscaling of the STE*

A major advantage of STEs over conventional THz sources is their straightforward scalability. Accordingly, an optimized large-area trilayer STE with a diameter of 7.5 cm was able to generate strong THz pulses, the peak electric field of which reached ~300 kVcm$^{-1}$ at a bandwidth of ~15 THz upon pumping with 60 fs pulses with an energy of 5 mJ and a center wavelength of 800 nm [69]. This THz electric field strength allowed one to measure the THz Kerr effect in diamond, which scales quadratically with the THz electric field. Future improvements in the STE THz-emission efficiency are, therefore, foreseen to enable studying THz strong-field phenomena over a wide THz frequency range [169].

2.3.10 *THz emission induced by other spintronic phenomena*

*Spin-to-charge current conversion inside the magnetic material.* Besides the ISHE mechanisms inside the NM, S2C can also occur already inside the magnetic material such as in single-layer FM samples [170-172] but also in FM|NM heterostructure. In single-layer FM samples, the out-of-plane spin current may arise from pump-induced gradients or structural gradients inside the sample. Accordingly, the measured THz-emission signals from MgO|FM|Quartz [FM = (Fe$_{0.8}$Mn$_{0.2}$)$_{0.67}$Pt$_{0.33}$, Fe$_{0.8}$Mn$_{0.2}$, Co$_{0.2}$Fe$_{0.6}$B$_{0.2}$, and Ni$_{0.8}$Fe$_{0.2}$] were shown to follow the same symmetries as the ISHE-based STE [Eq. (3a)]. However, the THz generation efficiencies for STEs based on S2C inside the FMare far below state-of-the art STEs based on the ISHE inside NMs such as Pt.

*Magnetic dipole radiation.* According to Eq. (2), a time-varying magnetization can emit a THz pulse [44]. Recent studies found that the corresponding magnetic-dipole radiation from metallic FM layers of a few nanometers thickness is on the order of 1% of the electric dipole radiation obtained from optimized STEs under identical excitation conditions [113]. However, care needs to be taken when assigning electric- or magnetic-dipole character to the observed THz-emission signal [44]. It requires a thorough symmetry analysis of the THz signal as shown recently [46, 113].

*Inverse spin-orbit torque.* Apart from the ISHE and the IREE, THz emission triggered by the inverse spin-orbit torque (ISOT) was reported in Co|Pt [173] and CoFeB|Ag|Bi [123] thin films under excitation with circularly-polarized femtosecond laser pulses as well as for NiO|Pt [139] thin films with linearly-polarized pump pulses. It was suggested that the optical spin transfer torque, which is also known as the resonant inverse Faraday effect, leads to a tilting of the in-plane magnetic order in the sample plane [48, 173].

For Co|Pt [173] and CoFeB|Ag|Bi [123], the subsequent magnetization dynamics injects a spin current polarized perpendicular to **M** into the adjacent NM, e.g., NM=Pt. The spin current is converted via the ISHE inside the NM into a charge current. This ISOT-based THz generation is phenomenologically described by [173]

$$\mathbf{E}_{\text{THz}} \propto \mathbf{j}_c \propto \mathbf{n} \times [\mathbf{M} \times \boldsymbol{\sigma}] I, \tag{9}$$

Where $\boldsymbol{\sigma}$ is the axial unit vector pointing parallel or antiparallel to the propagation of the circularly-polarized light, and $I$ is the pump-pulse intensity envelope. Importantly, the polarization of $\mathbf{E}_{\text{THz}}$ for the ISOT mechanism is oriented along **M** in the studied FMs, whereas it is perpendicular to **M** for the ISHE and IREE mechanisms [see Eq. (3a)].

Thus, upon illuminating a Co|Pt heterostructure with circularly polarized femtosecond laser pulses, two THz emission signals can be detected with different polarizations, perpendicular and parallel to **M** (Fig. 15a). The ISOT-component follows Eq. (9), where the sign change of **M** and $\boldsymbol{\sigma}$ leads to the reversal of the emitted THz waveform (Fig. 15b). In CoFeB|Ag|Bi, a similar dependence of the emitted THz signal was observed [123]. However, the measured THz-signal amplitudes from the ISOT effect strongly depend on the sample preparation processes, and the ISOT-based THz emission tends to decrease with FM/NM interface quality In terms of THz-generation efficiency, the ISOT-based structures were found to deliver THz amplitudes that were significantly smaller than from ISHE/IREE-based STEs [123, 173].

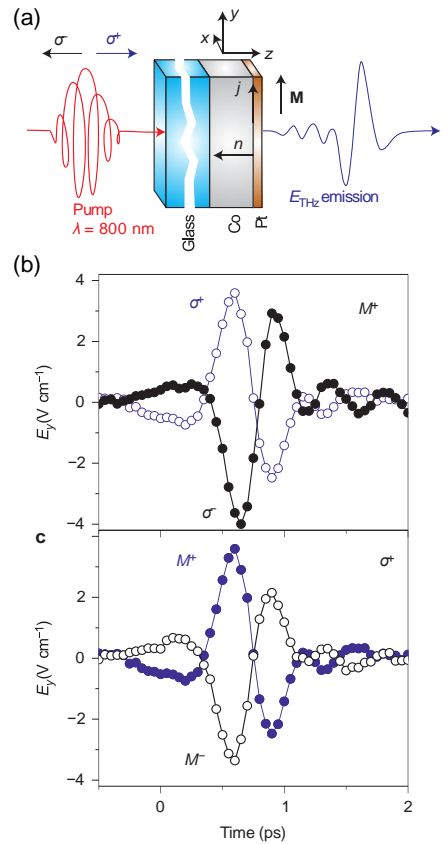

Figure 15. THz emission from Co|Pt based on the inverse spin-orbit torque. (a) Upon exciting a FM|NM thin film heterostructure with circularly-polarized optical femtosecond laser pulses, the optical spin transfer torque tilts the in-plane magnetization **M** of the FM layer. Subsequently, the magnetization relaxes back to its equilibrium orientation thereby emitting a spin current into the adjacent Pt layer. The inverse spin Hall effect converts the spin current into an in-plane charge current that radiates at THz frequencies. The polarization of the THz pulse is along **M**. (b) THz emission signals for the two pump helicities ($\sigma^{+,-}$) at constant **M**. (c) THz emission signal for opposite **M** at constant pump helicity. The figures are adapted from Ref. [173]. Reproduced with permission from Nat. Nanotech., 11, 455-8, 2016. Copyright 2016 Macmillan Publishers Limited.

As of now, none of the above alternative spintronic phenomena are competitive with respect to STEs based on the ISHE S2C.

**In conclusion, all these efforts demonstrate that novel designs of the STE can greatly enhance its efficiency as well as functionality, and promote this type of THz emitter toward real-world applications.**

3. **Current understanding of elementary processes in the STE**

The physical phenomena in STEs are manifold, including ultrafast spin dynamics, transport and S2C mechanisms. To improve the STE performance, the understanding of the microscopic mechanisms behind these different steps is mandatory, including a tailoring of the STE to specific applications. We focus on STEs that can be described according to the 4 processes introduced in Sects. 2.1 and 2.3.2. In the following, we address processes (1)-(3) in more detail.

### 3.1 *Processes (1), (2): Laser-driven spin-current generation*

At frequencies much smaller than ~1 THz, it is known that spin currents can be triggered by gradients of the electrostatic potential (leading to longitudinal spin-polarized electron currents in metallic ferromagnets or transverse spin currents by the ISHE in metals), by temperature gradients (e.g. leading to longitudinal spin-polarized electron currents in metals due to the spin-dependent Seebeck effect or magnon currents due to the spin Seebeck effect) and by gradients of the spin voltage [87]. The latter is also known as spin accumulation and quantifies the excess (or lack) of local spin density, i.e., how far the local spin density of a solid is away from its equilibrium value. Transferring these concepts to THz frequencies and nonthermal states is a matter of current research [88, 113, 174].

The spin-current generation and injection triggered by femtosecond laser pulses in FM|NM heterostructures is strongly related to the ultrafast spin and carrier dynamics in the FM layer. Usually, in a NM (e.g. Au or Ag), the ultrafast relaxation of photoexcited electrons happens on a sub-picosecond time scale, and the associated dynamics of energy transfer from the orbital degrees of freedom to the phonons can be described by the two-temperature model (2TM) [175, 176]. In the limit of weak excitation, the 2TM can be extended to nonthermal electron distributions [113, 174]. In a FM, however, ultrafast optical excitation additionally causes a magnetization quenching on a sub-picosecond timescale, as first observed by Beaurepaire *et al.* [177].

In thin single FM films on insulating substrates, such ultrafast demagnetization is governed by transfer of spin angular momentum from the electrons to the crystal lattice [44, 45, 177-198]. In FM films thicker than the penetration depth of the pump or in stacks such as FM|NM, an additional channel for the transfer of spin out of the excited FM regions becomes possible: ultrafast or THz spin transport as visualized in Fig. 16 [98, 199-208].

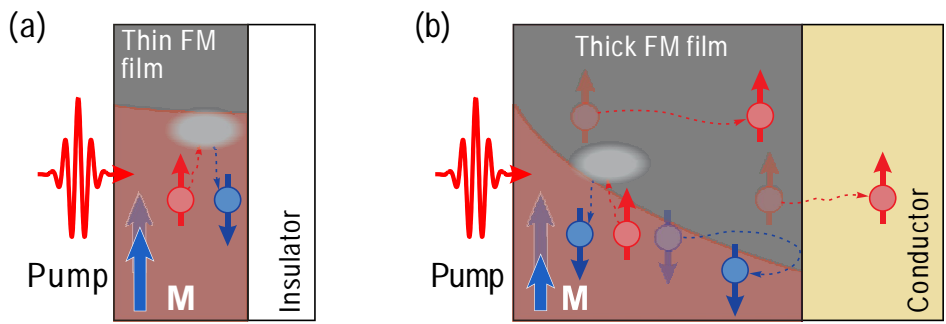

Figure 16. Schematics of femtosecond laser excitation of (a) a thin FM on an insulating substrate, allowing for local relaxation processes only, and (b) a thick FM on a conducting substrate, where both local and non-local relaxation processes are present. The red-shaded areas represent the in-depth profiles of the laser excitation. Reproduced with permission from J. Phys. Cond. Mat., 29, 174002, 2017. Copyright 2017 IOP Publishing Ltd.

### 3.1.1 *Superdiffusive spin transport*

To simulate the ultrafast electron spin current following ultrafast laser excitation, Battiato *et al.* developed a semiclassical model of spin transport [98, 199]. Their theory captures both ballistic transport (without electron scattering) and diffusive transport (dominated by electron scattering), which are summarized as superdiffusive spin transport (SDST). The simulations based on the SDST model are supported by different ultrafast optical studies [202-207], although some experimental results indicated that other mechanisms might significantly contribute, too [209-213].

In the SDST model, rate equations describe the spin-dependent carrier population, together with the *ab initio* input of the carrier velocities and lifetimes. Several events are taken into consideration, including multiple spin-conserving electron collisions and inelastic electronic scattering cascades [98]. In 3d transition FMs, the photoexcited sp electrons generally have higher band velocities and longer lifetimes than the excited minority spins. According to the SDST model, this effect leads to spin transport out of the FM layer upon optical excitation.

SDST simulations could successfully reproduce the qualitative shape of the spin-current dynamics and its order of magnitude in Fe|Au and Fe|Ru bilayers [63]. However, the connection of the SDST model to spintronic concepts such as gradients of spin voltage, temperature or electrostatic potential is not obvious and was addressed only recently [113].

In addition to the superdiffusion model, a particle-in-cell simulation based on Boltzmann equation was developed recently [214]. It can reproduce the superdiffusive transport behavior and also captures the main physical process of the laser-induced demagnetization.

### 3.1.2 *Optically induced intersite spin transfer*

The term optically induced intersite spin transfer (OISTR) refers to a coherent, ultrafast spin transfer between two subsystems that occurs while the pump laser pulse overlaps with the sample [215]. It was predicted by ab-initio calculations and experimentally observed in Ni|Pt bilayers [216]. As typical spin-current dynamics in STEs were found to strongly exceed the pump pulse duration [113, 114, 116], OISTR is considered to make a minor contribution in the THz-emission process. However, the experimental observation of the OISTR effect using THz spectroscopy poses a highly interesting challenge for future studies.

### 3.1.3 *Spin voltage as a major spin-current driving force*

In general, three different forces can drive a spin current [87, 217]: Gradients of the electrostatic potential, of the temperature or of the spin voltage. The latter is also known as the spin accumulation and quantifies how much the local spin density deviates from its instantaneous equilibrium value. A transient spin voltage was experimentally observed in laser-excited single Fe films [218]. Moreover, theoretical works further suggest that the spin voltage has a major impact on driving ultrafast demagnetization in FMs [89, 219, 220].

Recently, Rouzegar *et al.* [113] made an experimental and theoretical effort to connect ultrafast spin transport and ultrafast demagnetization to differences or gradients of macroscopic parameters such as temperature and spin voltage. To this end, they used Boltzmann-type rate equations and the Stoner model, which describes the electronic structure of a typical 3d FM qualitatively well. The magnetization **M** shifts the energy of spin-up and spin-down electronic Bloch states by an energy proportional to $|\mathbf{M}|$ with respect to each other (see Fig. 17).

When such a FM is excited by an ultrafast optical pulse, the chemical potential of each spin channel changes differently. As a result, a splitting of the spin chemical potential, an ultrafast spin voltage, arises [89, 217, 219]. The relaxation of this spin voltage, which causes a magnetization quenching, can occur via local spin-flip scattering events inside the FM or via spin transport into an adjacent layer (Ref. [113] and Fig. 17), which is also consistent with observations in other THz emission studies [221]. It is noteworthy that the concepts of temperature, chemical potential and, thus, spin voltage can be extended to nonthermal electron distributions that prevail directly after optical excitation [113, 222].

Experimental evidence that spin currents and demagnetization evolve with the same ultrafast dynamics in FM|NM heterostructures was found directly in THz emission studies comparing FM and FM|NM samples [113] or by probing the magnetization dynamics via the magneto-optical Kerr effect (MOKE) in more complex magnetic multilayers [223, 224]. An important insight from the results in Ref. [113] is that the ultrafast spin current in the STE is only determined by the amount of excess energy that is deposited in the electronic system. In other words, only an ultrafast heating of the FM is required, regardless of the precise shape of the resulting nonequilibrium electron distribution and, thus, the size of the pump photon energy. This notion is in line with previous experimental findings for infrared and visible [67, 68, 157] down to THz pump photons [174].

In terms of the STE performance, the connection between ultrafast demagnetization and spin transport can be of great value. It means that the vast knowledge from the ultrafast-magnetism community [84] can be transferred and exploited to enhance the THz-emission strength of the STE. A possible route to an enhanced STE performance would be the suppression of local spin relaxation inside the FM and the corresponding enhancement of spin transport.

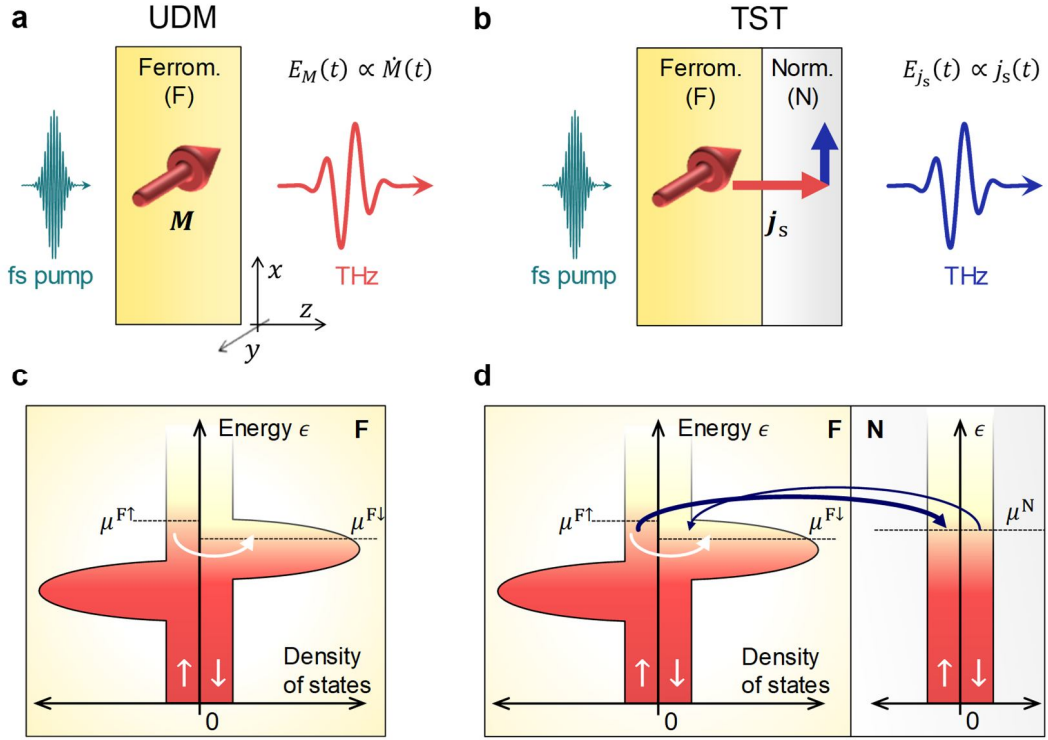

Figure 17. Comparison of (a) optically driven ultrafast demagnetization (UDM) and (b) THz spin transport (TST). While the former causes magnetic dipole radiation proportional to the magnetization dynamics $\dot{M}(t)$, the latter drives electric dipole radiation that is usually found in the STE. (c, d) Excitation by a femtosecond optical laser pulse triggers the splitting of the spin chemical potentials $\mu^{F\uparrow}$ and $\mu^{F\downarrow}$ inside the ferromagnet. This ultrafast spin voltage can relax via (c) local spin-flip events or via (d) spin transport into an adjacent nonmagnetic material. The figure is adapted from Ref. [113].

3.1.4  *Spin injection across the FM/NM interface*

As shown in Fig. 2, the spin current inside the STE is strongly localized at the FM/NM interface [115], which is why, in addition to the bulk, the interfaces have a strong impact on the spatial distribution of $\mathbf{j}_s$ [171]. For simplicity, theoretical calculations often assume a transparent FM/NM interface for electron or spin transport, i.e., all electrons can pass through the interface without any loss. However, in reality, electron reflection and spin loss at the interface have to be considered and can be seen as an interfacial spin resistance.

Recently, a theoretical study calculated the reflection of a superdiffusive spin current inside a FM|NM heterostructure (see Fig. 18) and revealed a possible impact of the interfacial spin resistance on ultrafast spin currents [90]. The calculated results show that the spin-up electrons (majority) have a lower reflectivity, that is, a higher interface transmittance than the spin-down electrons (minority) at an Fe/NM interface. Accordingly, the authors of Ref. [90] argued that the spin current would have a higher spin polarization after crossing the interface. Such a "positive" spin filter effect might enhance the demagnetization in the FM layer and the THz-emission efficiency. In the case of the Ni/NM interface, the spin filter effect is of opposite ("negative") character and would lead to a decrease in demagnetization and THz-emission amplitude. [90] would be reduced.

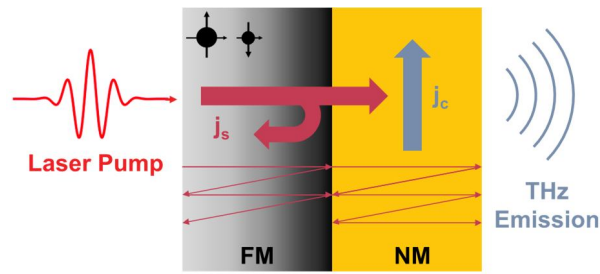

Figure 18. Ultrafast spin transport and THz emission in the FM|NM heterostructure including the interface reflection. Wen-Tian Lu, Yawen Zhao, Marco Battiato, Yizheng Wu, and Zhe Yuan, Phys. Rev. B 101, 014435, 2020; licensed under a Creative Commons Attribution (CC BY) license.

A well-established method for DC characterization of spintronic properties of FM|NM heterostructures is spin pumping [225]. However, the quantity that determines the spin injection into the NM layer across the interface in spin-pumping experiments is the effective spin mixing conductance $g_{\text{eff}}^{\uparrow\downarrow}$, which is different from the diagonal counterparts $t_{\text{FM/NM}}$ Accordingly, a recent study, which directly measured $g_{\text{eff}}^{\uparrow\downarrow}$ and $t_{\text{FM/NM}}$ in the same samples by spin pumping and THz-emission spectroscopy [226], respectively, found that these two quantities are different but can be related to one another if the spin loss by interfacial spin-orbit coupling, that is, a spin memory loss during the THz emission process [227], is included. The spin memory loss, however, needs to be characterized by independent experiments such as tr-MOKE measurements [205]. The experimental data suggested that, in Ni|Au heterostructures, ~50% of the spin current is lost in the spin transfer process at the interface due to spin-flip scattering processes.

Indeed, a recent work [100] (see Sect. 2.4.3) further confirmed that the quality of the interface plays an important role in the spin injection process, and that the interfacial disorder can result in a reduction of the spin current from FM to NM. Future interface-modification studies might aim at disentangling the role played by changes in $t_{\text{FM/NM}}$, S2C by interfacial alloys [115] or a change in the interfacial magnetic properties such as reduced Curie temperatures [228, 229] that would impact the spin voltage [113] close to the interface.

### 3.2 *Process (3): Spin-to-charge current conversion*

S2C plays an important role in the THz emitter made of FM|NM heterostructures. S2C phenomena such as ISHE and IREE are a consequence of SOC mainly in the NM. Therefore, materials, such as heavy metals, topological insulators, two-dimensional transition metal dichalcogenide (2D-TMDCs) and Weyl semimetals, can all be candidates for the NM layer. This flexibility, thus, enables to probe spin-related phenomena in a wide range of materials beyond purely metallic FM|NM heterostructures using THz-emission spectroscopy.

#### 3.2.1 *Spin Hall effect and inverse spin Hall effect*

The SHE is a transport phenomenon that generates spin currents from charge currents via SOC [230-233]. The inverse process, the ISHE, which converts a spin current into an electric current, is driven by the same mechanism: a SOC-induced spin-dependent scattering (Fig. 19). Different from the extrinsic SHE that originates from impurity scattering, the intrinsic SHE is directly related to the band structure of the material and has a sizeable S2C efficiency in heavy metals (e.g. Pt, W and Ta) and topologically nontrivial materials (e.g. topological insulators and Weyl semimetals).

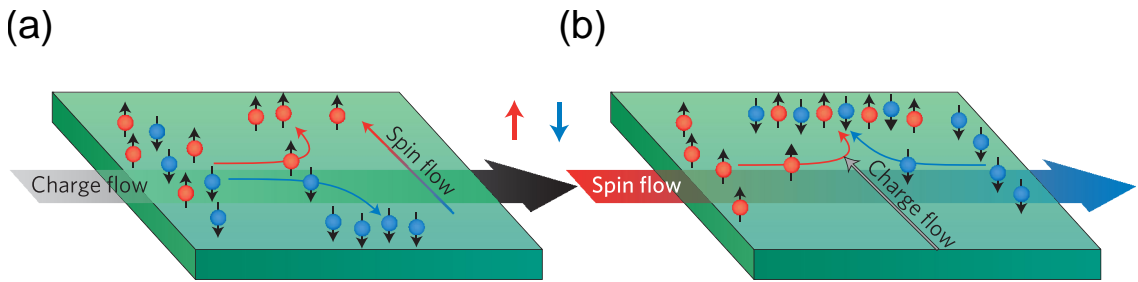

Figure 19. Schematic of (a) spin Hall effect and (b) inverse spin Hall effect. Reproduced with permission from Nat. Phys., 10, 549-550, 2014. Copyright 2014 Macmillan Publishers Limited.

The relation between charge- and spin-current density $j_c$ and $j_s$ is given by Eq. (3a), which is repeated here for convenience [231]:

$$j_s = \gamma j_c. \tag{10}$$

The parameter $\gamma$ is the spin Hall angle that describes the S2C efficiency [see Eq. (3a)]. In some heavy metals, such as Pt and Ta, $\gamma$ reaches values between $10^{-2}$ to $10^{-1}$ [234-239], which is much larger than for materials with minor SOC (Cu, Au or some semiconductors) [240, 241]. Up to date, there is a huge number of DC spintronic studies of the SHE/ISHE in various sample system that are indispensable for realizing efficient STEs.

Aside from $\gamma$, another critical parameter in the SHE/ISHE is the relaxation length $\lambda_{NM}$, which describes how far the spin current can flow in a material. At THz frequencies, it equals the electronic mean free path rather than the spin diffusion length, as discussed in Sect. 2.3.4.

3.2.2 *Rashba-Edelstein effect and inverse Rashba-Edelstein effect*

Different from the SHE and ISHE, which are allowed in regions with inversion symmetry, additional S2C mechanisms can occur in regions with broken inversion symmetry such as interfaces or surfaces. Prominent examples are the Rashba-Edelstein effect (REE) and its inverse, the IREE (Fig. 20). In a simple microscopic model of the REE, an applied electric field drives a charge current and, thus, shifts the Fermi surface, that is, it induces an asymmetry in the electron distribution in wavevector space (Fig. 20b). Due to spin-momentum locking, such a nonequilibrium electron distribution can lead to an asymmetric spin distribution, and, thus, form a net spin polarization in regions with broken inversion symmetry. The spin polarization can diffuse as a spin current or exert spin torque to an adjacent material [242].

REE and IREE were observed experimentally in many material systems, such as at Ag/Bi interfaces [117, 123, 126, 243-246], in topological insulators [125, 247-252], two-dimensional electron gases (2DEGs) [253], 2D-TMDCs [254-256], Heusler compounds [257], and at metal/oxide interfaces [258].

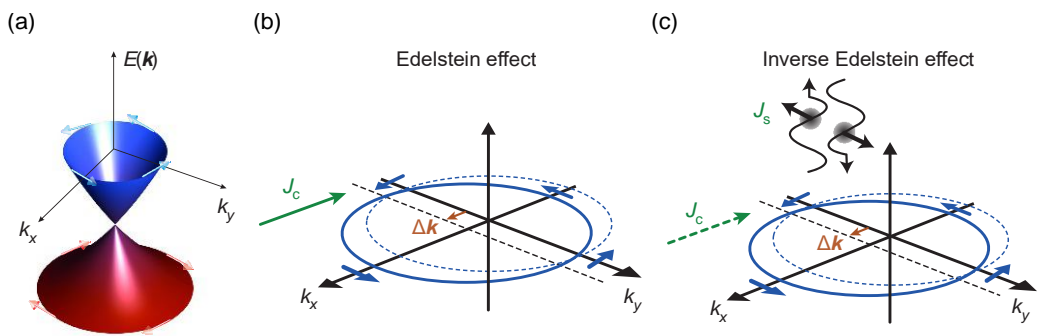

Figure 20. (a) Schematic of band structure of the surface states of a topological insulator. Schematic of (b) Rashba-Edelstein effect and (c) Inverse Rashba-Edelstein effect. Reproduced with permission from Nature, 539, 509, 2016. Copyright 2016 Macmillan Publishers Limited.

The REE and the IREE can be characterized by the following S2C coefficients:

$$q_{\text{REE}} = \frac{j_s}{J_c}, \tag{11}$$

$$\lambda_{\text{IREE}} = \frac{J_c}{j_s}, \tag{12}$$

where $j_s$ is a 3D spin current with dimension s$^{-1}$m$^{-2}$ and $J_c$ is a 2D charge current with dimension s$^{-1}$m$^{-1}$. Therefore, the conversion coefficients $q_{\text{REE}}$ and $\lambda_{\text{IREE}}$ have dimensions m$^{-1}$ and m, respectively. In addition, $q_{\text{REE}}$ and $\lambda_{\text{IREE}}$ are constants that depend on the properties of the interface/surface [126, 248, 251, 259].

To compare the REE/ IREE with the SHE/ISHE, a parameter called effective spin Hall angle or spin-torque ratio, $\gamma^{\text{eff}}$, is frequently used. This parameter can be defined as $\gamma^{\text{eff}} = \lambda_{\text{IREE}} t$ for the IREE and $\gamma^{\text{eff}} = q_{\text{REE}}/t$ for the REE, respectively. Here, $t$ is the thickness of the interface or surface. Although the thickness of a 2D interface/surface is in principle hard to define, $t$ is often approximated as the thickness of the first few atomic layers, in which the S2C happens. For example, in the spin-pumping experiments of 4-layer MoS$_2$ [255], the IREE coefficient $\lambda_{\text{IREE}} = 0.4$ nm and $\gamma^{\text{eff}} = 0.96$ were found by taking the thickness of the interface equal to the sample thickness, that is, $t = 2.4$ nm. Up to date, the materials that show a sizeable REE/IREE usually have extremely high S2C efficiencies, which raises questions about the exact meaning of $\lambda_{\text{IREE}}$. For instance, the value of $\gamma^{\text{eff}}$ is between 0.14 and 0.96 for MoS$_2$ [254, 255], between 2.0 and 3.5 for Bi$_2$Se$_3$ [234, 247, 249], and ~1.5 for a Ag/Bi interface [126].

In principle, distinguishing the ISHE from the IREE can be possible by the different thickness dependences of the induced charge currents that are given by [126]:

$$J_c = e\gamma \lambda_{\text{NM}} \tanh \frac{d_{\text{NM}}}{2\lambda_{\text{NM}}} j_s \tag{13}$$

in the ISHE case (see also Eq. (6)) and by

$$J_c = e\lambda_{\text{IREE}} j_s \tag{14}$$

in the IREE case. However, an experimental separation of the ISHE from the IREE remains challenging in transport [260, 261] and especially in THz-emission experiments. The latter might be eased in the future by the observation that in contrast to the ISHE, the IREE scales with the accumulated spin density and should, therefore, exhibit memory effects, which imply a frequency dependence of $\lambda_{\text{IREE}}$ and $q_{\text{REE}}$.

Experimentally, strong evidence for the IREE mechanism was provided by the opposite polarities of the emitted THz pulses depending on the stacking order, i.e., by comparing

Ag/Bi to Bi/Ag interfaces (Fig. 21) although certain ambiguities remain [262]. The bandwidth of the detected THz pulses was about 3 THz and limited by the relatively narrow bandwidth of the detection crystal [117, 123]. It appears reasonable to the authors that the IREE-based STEs can reach the record bandwidths of their ISHE-based counterparts given the observation that many other spintronic phenomena also remain operative at THz rates [64, 78, 142, 263].

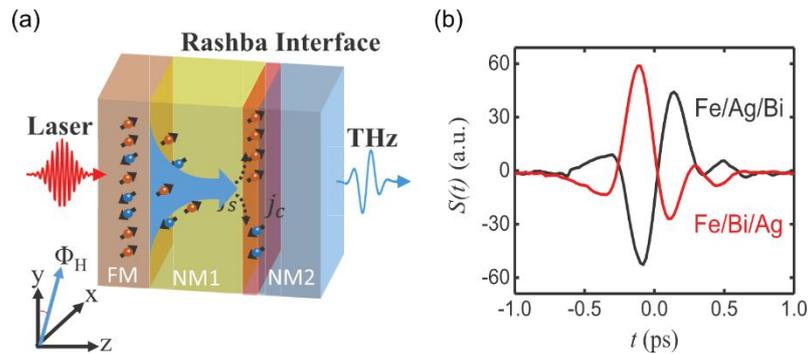

Figure 21. THz emission from Ag/Bi Rashba interfaces under femtosecond-laser excitation. (a) Excitation of a thin-film heterostructure with a femtosecond laser pulse triggers the injection of a spin current into the NMs (NM1=Ag, NM2=Bi). The spin current is transformed into an in-plane charge current by the Rashba spin-orbit field at the Ag/Bi interface. (b)The polarity of the emitted THz pulse depends on the orientation of the Rashba interface with respect to the spin current source (FM layer). Reproduced with permission from Phys. Rev. Lett., 121, p. 086801, 2018. Copyright 2018 American Physical Society.

Another example of IREE-based THz emission appears in Co|$MoS_2$ heterostructures [124]. Previous spin-torque-ferromagnetic-resonance and spin-pumping measurements revealed a sizeable S2C via the IREE in the atomically-thin 2D system $MoS_2$ [254-256]. Note that $MoS_2$ is a semiconductor with a large bandgap of ~1.9 eV, which is quite different from the conducting NM layers discussed before. Such material difference may lead to a much smaller spin-current-injection efficiency [264] but could also enable new functionalities, as will be discussed further below [265].

Remarkably, the THz-emission amplitude from IREE-based STEs can reach up to 20% of a 1-mm-thick ZnTe crystal under the same experimental conditions [117, 125] and, thus, also about 20% of the best ISHE-based STEs.

The high S2C efficiency of the IREE-based STE can potentially generate intense THz radiation and provide a new approach for broadband THz emission.

### 3.2.3 *Other S2C mechanisms*

Apart from the ISHE and the IREE, there are other S2C mechanisms, such as the valley Hall effect (VHE) [266] and valley Edelstein effect (VEE) [267]. Both mechanisms are associated with the valley degree of freedom that is coupled with the spin degree of freedom for instance in 2D transition metal dichalcogenides [268].

For example, a monolayer of $MoS_2$ has two non-equivalent valleys in its band structure that are located in the vicinity of the K and K' point. Due to spin-valley coupling, a spin flip is always accompanied by a change of the electronic valley and vice versa [268]. Therefore, the VHE, which separates charges transversely to an applied electric field, can be accompanied by the formation of a transverse spin accumulation. In contrast, for the VEE, the electric-field-induced spin polarization is predicted to be parallel to the applied electric field. However, the reciprocal effects of VHE and VEE, which would be useful for the STE design, have not yet been observed. Its observation would require the injection of a spin current into the VHE/VEE material and the separation from other S2C mechanisms such as the SHE or REE. Another challenge in terms of strong THz emission might arise from the difficulty in efficiently injecting spin currents from a magnetic metal into the typically semiconducting VHE/VEE material [264].

In summary, S2C is a critical ingredient for spintronic THz emission and has a strong impact on the THz-emission efficiency. Conversely, THz-emission spectroscopy is an excellent tool itself to unravel the properties of S2C in various material systems [82]. It is, therefore, a promising approach to find novel efficient spintronic materials.

## 4. Applications

The unique advantages of the STE over other THz emitters enable a manifold of applications. For each of them, different aspects of the STE are exploited and, thus, require a tailored optimization that may be guided by our introduced STE model (Eqs. (3a), (3b), (3c) and (6)). In the following, we will highlight certain emerging STE applications that may have a profound impact beyond the field of THz photonics.

### 4.1 *Broadband linear THz spectroscopy*

In several recent studies, STEs were successfully used to perform broadband THz spectroscopy. In an early work, Seifert et al. demonstrated the spectroscopic characterization of a thin Teflon tape from 1 to 30 THz, which could reveal the characteristic THz phonon absorptions of this material [64]. Subsequent studies used STEs and linear THz spectroscopy to measure the frequency dependence of central spintronic phenomena such as the anomalous Hall effect or the anisotropic magnetoresistance [79, 263].

Davies et al. used a STE to measure the transient photoconductivity over an extended THz spectral range in perovskites [269] and to determine the temperature-dependent THz refractive index of quartz [270]. In a related study, the equilibrium and the transient THz conductivity of a perovskite sample as a function of doping concentration was measured with a STE [271].

In a near-field-type approach, Balos and coworkers measured the THz complex-valued refractive index of water that was in direct physical contact to the STE's metal surface from 0.3 to 15 THz [272]. This technique can be applied to any liquid and is particularly useful for THz-opaque liquids.

4.2 *Spatiotemporal THz beam modulation*

One major advantage of the STE over other THz sources is its tunability by external magnetic fields, which allows for the global variation or even spatial patterning of the linear THz polarization.

Hibberd and coworkers [76] demonstrated reversible magnetic patterning of the FM layer inside the STE using inhomogeneous external magnetic fields as shown in Fig. 22a. This approach permits the creation of new THz polarization states, including doughnut-like modes or radially polarized THz electric fields [76, 273]. A related study generated THz pulses from a STE that had two perpendicularly magnetized regions close to each other [110]. This approach allowed for the creation of elliptically polarized THz electric fields. Because it is based on double pulses, it is applicable to a relatively narrow frequency range.

Shaping the THz polarization from linear to circular behind the STE was demonstrated with the help of a large-birefringence liquid crystal, onto which the STE was deposited [110]. Another approach exploits different THz-emission mechanisms in STEs including topological materials. It superimposes THz electric fields with perpendicular linear polarization, yet with different phases to obtain elliptically polarized THz pulses over a relatively narrow THz frequency interval [274]. Future STE designs might additionally incorporate multilayer structures with a pinned FM layer to allow for a magnetic-field-free operation [107] as required in many application environments.

Recently, Gueckstock et al. succeeded in modulating the THz pulses with a bandwidth of 30 THz in polarity and polarization direction at rates of up to 10 kHz and with a contrast exceeding 99%. This achievement relies on the rapid variation of the STE magnetization as shown in Fig. 22b [75, 275]. It enables low-noise broadband THz spectroscopy or rapid ellipsometric sample characterization, without the need for mechanical choppers or other modulators of the optical pump beam.

Finally, shaping of THz beam properties such as the divergence can be achieved by depositing the STE on flexible substrates as shown in Fig. 22c [66]. A straightforward extension of this approach involves growing the STE on top of curved surfaces. It allows for a built-in THz focusing functionality by using for instance parabolic mirrors or THz/optical lenses as the STE substrate.

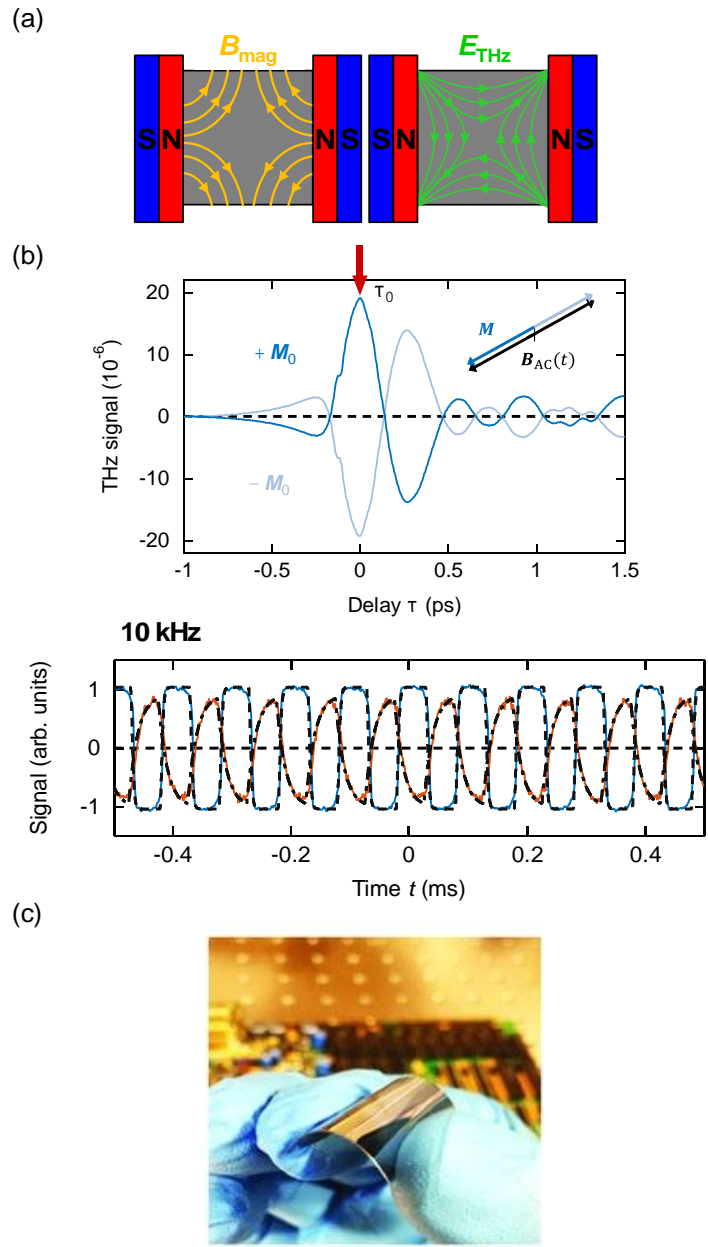

Figure 22. Spatiotemporal THz beam shaping using the STE. (a) Spatially structured magnetic fields can generate doughnut-like THz-electric-field distributions from the STE. Reproduced from Appl. Phys. Lett. 114, 031101, 2019, with the permission of AIP Publishing. (b) Modulation of the STE magnetization at kHz rates enables rapid THz-polarity modulation. Left panel: THz-emission signals for opposite magnetizations of the STE. Right panel: Time evolution of the peak THz signal (red arrow in left panel) upon modulation of the STE magnetization at 10 kHz (blue line) and magneto-optic signal tracking the STE magnetization (orange line). Gueckstock, O., L. Nadvornik, T.S. Seifert, et al., Optica, 8, 1013, 2021; licensed under a Creative Commons

Attribution (CC BY) license. (c) Flexible substrates or deposition on curved surface of STE stacks broaden their applicability. Reproduced with permission from Adv. Mat., 29, 1603031, 2017. Copyright 2017 Wiley-VCH.

4.3 *THz imaging*

The STE design provides a unique platform for near-field imaging applications because the THz emitter is just a few nanometers thick. Therefore, probing THz near-fields with lateral features above the STE surface becomes possible whose size is given by the lateral structure of the optical pump beam that is incident onto the STE. In principle, the lateral THz field can have feature sizes comparable to the wavelength of the optical pump, which often lies in the submicrometer range [276].

Recently, such THz near-field imaging with a STE was first demonstrated [77], which allowed the authors to resolve subwavelength structures with a resolution of 6.5 µm by placing the object in the near-field region of the STE. In their experiment, the STE was illuminated with a spatially structured pump beam, and the emitted THz signal was recorded using EO detection. After illumination with a specific set of spatially structured pump pulses, an inversion algorithm served to extract the image of the object. Polarization-induced imaging artifacts could be reduced by switching the THz polarization with an external magnetic field. Depth information of the object could be obtained by exploiting the different arrival times of the THz pulses at the detector.

A similar approach aimed at combining the STE near-field imaging concept with resonant microstructures, i.e., split ring resonators, to enhance certain THz frequencies locally. In this way, Bai and coworkers could detect a change in the emitted THz pulse upon covering their STE device with cancer cells, which might be useful for future biological sensing applications [277]. Recent studies reported on the implementation of a trilayer STE into a THz-emission microscope [278] and on achieving a spatial resolution down to 5 µm using THz near-field imaging with a STE [279].

Because of its compactness, a STE was successfully combined with an electro-optic detection unit to yield a magnetic field sensor with a relatively small footprint [280]. Following this approach, Bulgarevich et al. demonstrated a sensitivity in the millitesla range with submillimeter spatial resolution.

4.4 *Nonlinear THz spectroscopy*

Strong THz pulses can trigger a nonlinear response of the studied system [42]. As illustrated in Fig. 23 (a)-(b), a large-area STE was used to generate THz electric fields with a peak strength of 300 kV/cm$^{-1}$ [69] in the diffraction-limited focus of a Gaussian beam. The pulses drove a third-order nonlinear effect, i.e., the THz Kerr effect, inside a diamond sample.

A completely different nonlinear application of the STE was demonstrated by Müller et al. [281]. They coupled THz pulses with a bandwidth exceeding 20 THz generated from a STE by using 8 fs pump pulses into the junction of a scanning tunneling microscope (STM) as shown in Fig. 23(c)-(d). Because of the inherent nonlinear response of the tunnel current to the incident THz electric field, the THz pulses were rectified. The resulting DC current was detected by the low-bandwidth STM electronics. The broad THz spectrum delivered by the STE allowed the authors to extract the THz response function of the tunnel junction in the range between 1 and 15 THz, which revealed a pronounced low-pass behavior.

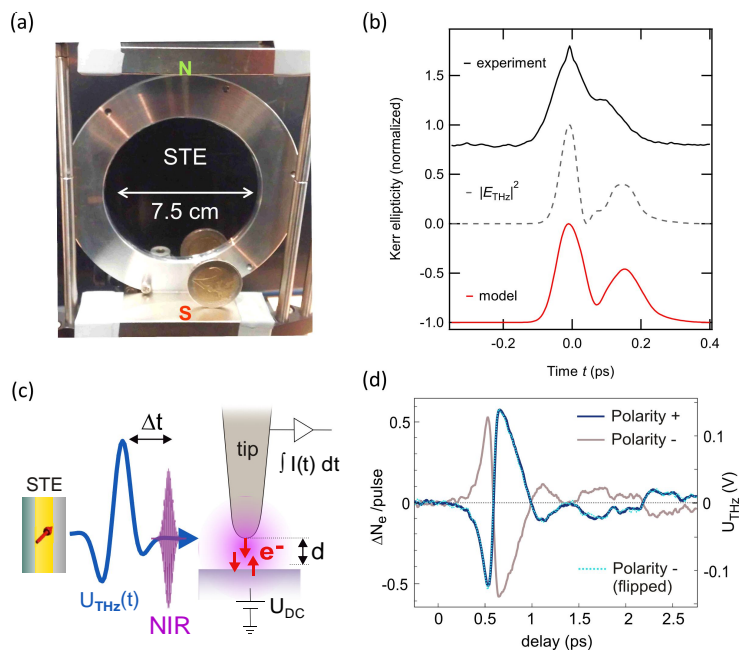

Figure 23. Nonlinear-optics applications of STEs. (a) Large-area STE emitter that can generate THz electric field strengths of 300 kVcm$^{-1}$ when excited by femtosecond laser pulses at 800 nm central wavelength and 5 mJ pulse energy. (b) Third-order nonlinear effect (Kerr effect) in diamond driven by the strong THz electric field generated with the STE shown in (a). Reproduced from Appl. Phys. Lett. 110, 252402, 2017, with the permission of AIP Publishing. THz-STM setup, in which the STE generates THz-electric-field pulses that serve to apply an ultrafast voltage across the STM junction. The resulting THz voltage $U_{THz}$ drives photoelectrons that are generated by a preceeding near-infrared (NIR) pulse. (d) Rectified tunneling electrons as a function of delay time between the THz and the NIR pulse. The resulting waveform is odd in STE magnetization and low-pass filtered by the response function of the STM junction. Muller, M., N. Martin Sabanes, T. Kampfrath, et al., ACS Phot., 7, 2046, 2020; licensed under a Creative Commons Attribution (CC BY) license.

### 4.5 *THz-emission spectroscopy of spintronic transport*

Note that the spintronic-THz-emission process itself is nonlinear in the pumping field and, thus, a nonlinear spectroscopic approach. It was used to characterize spintronic transport in FM|NM STEs as a function of material composition. Such THz emission studies were shown to be fully consistent with low-frequency electronic characterization approaches in terms of extracted S2C efficiencies [78, 142].

Importantly, however, the all-optical approach does not require any sample microstructuring and, thus, allows one to extract spintronically relevant parameters such as the S2C efficiency and the spin-current propagation length with large sample throughput [73, 88, 105, 117, 124, 125, 133-136, 140, 142, 159, 173, 282].

A further interesting approach along these lines is to study the propagation of spin currents through interlayers [66, 133, 283] that might even be magnetically ordered [284, 285] or of the FM magnetization dynamics [159, 286] that might even involve its laser-induced reversal [287].

Recent THz-emission studies provided new insights into the magnetic structure of the ferrimagnetic half-metal $Fe_3O_4$ [114] and revealed the initial formation dynamics of the ultrafast spin Seebeck effect ([88] and Sect. 2.3.3).

## 5. **Future perspectives**

### 5.1 *Semiconductors and 2D materials*

In contrast to metal-based STEs, ultrafast spin injection (and THz emission) from a FM into a semiconducting layer (SC) can be influenced by an interfacial Schottky barrier that may reduce the spin injection efficiency, simultaneously but increase the spin polarization of the injected carriers (Fig. 24 and Refs. [264, 265, 288, 289]). Moreover, the propagation length of spin currents inside SCs can be significantly larger than in metals [290], which could boost the STE THz-emission performance. Consequently, FM|SC bilayers are a promising platform for efficient STEs.

2D materials are a novel material class with promising properties such as spin-valley locking or large S2C efficiencies [124, 291-295] for their use in STEs. Recently, a giant ultrafast spin injection into a SC was experimentally confirmed in FM|$MoS_2$ heterostructures [124, 254], where $MoS_2$ is an atomically thin SC with a bandgap of round 1.9 eV and strong SOC [296]. The latter property is essential for a large S2C efficiency that leads to the emission of THz radiation. The optically-injected spin-current density was estimated to be $4 \times (10^6 - 10^8)$ A cm$^{-2}$ [124], which is several orders of magnitude larger than the values obtained by other methods, such as spin pumping ($10^2 - 10^3$ A cm$^{-2}$) [297] and spin-transfer torque in magnetic tunnel junctions ($10^{-1} - 10^0$ A cm$^{-2}$) [298, 299] .

Future engineering of the Schottky barrier and the SC bandgap could optimize optically injected spin currents. Along these lines, Vetter et al. [300] investigated the spintronic THz emission from NiFe|n-GaN heterostructures. They found a significant change of the emitted THz amplitude as a function of the GaN doping level. The more metallic the SC was, the smaller the emitted THz pulse became. However, disentangling the spin currents injected into the SC layer from other effects such as a decrease of the sample impedance with increased doping level [last factor in Eq. (6)] or a S2C already inside the metallic FM layer [171] remains a challenge.

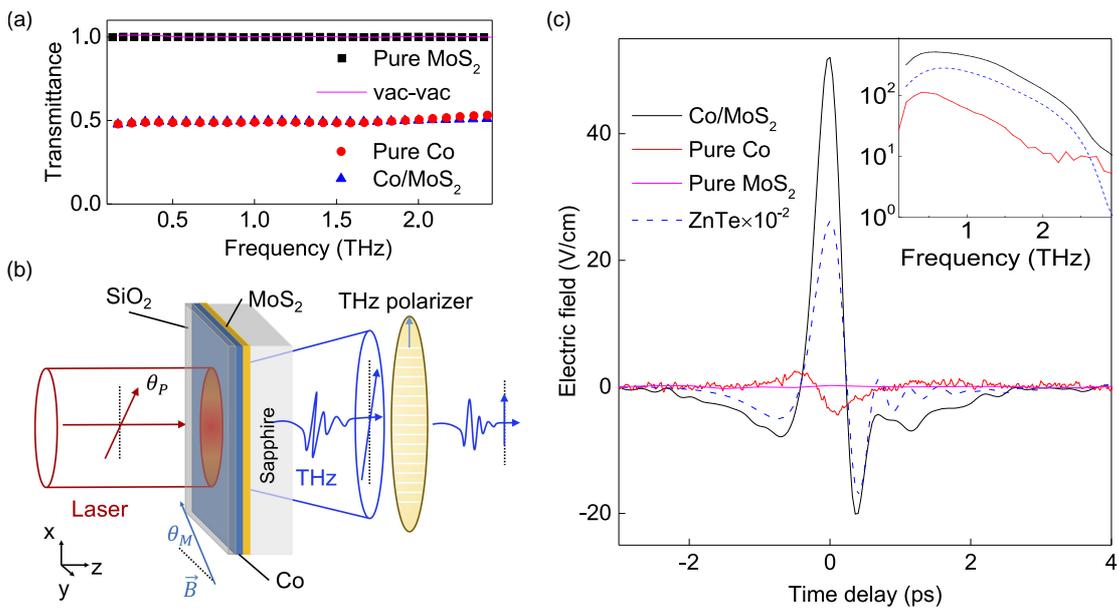

Figure 24. Measured spin injection from the FM Co into the semiconductor $MoS_2$. (a) THz transmittance through vacuum, Co, $MoS_2$ and Co|$MoS_2$ showing the transparency of $MoS_2$ to THz radiation. (b) Experimental geometry for measuring the THz emission upon optical pumping. (c) Comparison of emitted THz electric fields showing the enhanced emission from Co|$MoS_2$ bilayers due to spin injection into the semiconductor. Reproduced with permission from Nat. Phys., 15, 347-351, 2019. Copyright 2019 Macmillan Publishers Limited.

5.2 *Topological materials*

Topological materials bear a large potential for an efficient S2C due to the inherent spin-momentum locking realized for instance in the surface states of topological insulators. The impact of topological surface states on the THz S2C was investigated in Co|$Bi_2Se_3$ heterostructures [125]. As shown in Fig. 25, the THz-emission signal strongly depended on the $Bi_2Se_3$ thickness: The signal increased significantly if the $Bi_2Se_3$ layer was thicker than 5 quintuple-layers. This thickness is known to be the lower limit for the formation of the topological surface states [301]. However, a thorough distinction of the impact of the IREE and the topological surface state on the THz emission process poses an interesting question for future studies.

Interestingly, the THz-emission amplitude from Co|Bi$_2$Se$_3$ was found to be constant for a sample temperature between 10 and 300 K, which may make this type of emitter design suitable for applications in environments with large temperature variations such as space missions (Fig. 25b). Implementing emerging material classes such as magnetic or nonmagnetic Weyl semimetals [302] as S2C converters or spin-current-generating layers might in the future further enrich the STE by topological functionalities such as quantized photocurrents [303]. On the other hand, a recent study highlighted the efficient spin-current generation in Weyl semimetals that might help to enrich the STE by topological functionalities [304].

Moreover, the combination of the ISHE-based THz-emission mechanism with the circular photogalvanic effect, that is, the injection current, in topological materials was shown to allow for elliptical THz polarization states [305].

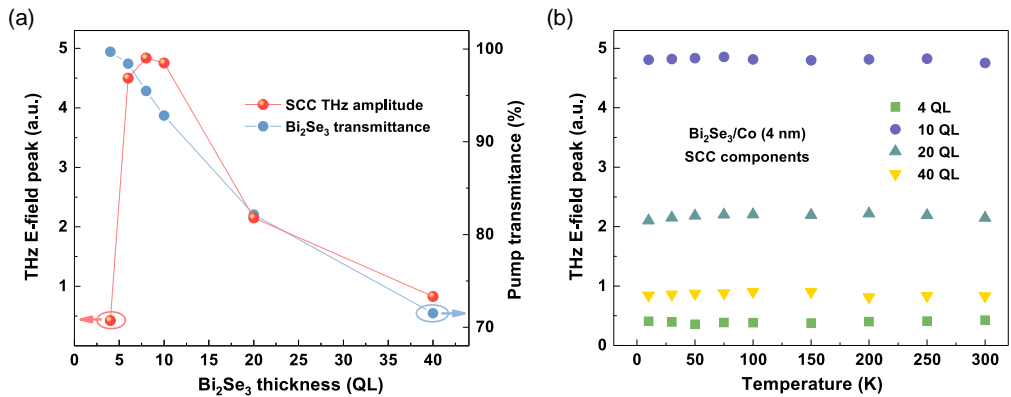

Figure 25. THz emission from thin film heterostructures containing Co and the topological insulator Bi$_2$Se$_3$. (a) THz-signal amplitude and THz transmittance as a function of the thickness of Bi$_2$Se$_3$. (b) THz-signal amplitude related to S2C as a function of temperature for different thicknesses, i.e., number of quintuple layers (QL). Reproduced with permission from Adv. Mat., 30, 1802356, 2018. Copyright 2018 Wiley-VCH.

5.3 *Photonic optimization*

Optimizing the STE from a photonic perspective mainly follows two goals: To increase (i) the pump-pulse absorption and (ii) the THz-outcoupling efficiency into free space.

Anti-reflective coatings were shown to be a promising approach to reach (i). Herapath et al. [67] could enhance the pump absorption from about 50% to 100%, thereby increasing the emitted THz pulse amplitude by a factor of 2 for certain pump wavelengths.

To reach goal (ii), THz antenna structures were deposited onto the STE by using patterned gold overlayers. As a result, the THz electric field from the STE was enhanced by a factor of 2 compared to an identical bare STE [108], as confirmed by a more recent study [306].

In another work, the STE performance was significantly improved in a narrow THz frequency range by adding a horn antenna to the STE [307].

It should, however, be noted that antenna designs and antireflection coatings typically only perform well in a certain interval of THz and pump frequencies, respectively. Thus, finding broadband solutions remains a future challenge, and there is much optimization potential for THz-antenna designs. An interesting strategy might be to guide the pump field to spatial regions, such as the FM layer, that enable larger THz-emission amplitudes per pump energy, for instance by plasmonic enhancement as used in PCS designs [57, 308].

5.4 *Integration, combination with existing technologies and readiness for real-world applications*

Hybrid emitter concepts hold great promise for combining the benefits from different THz-emitter designs. Chen and coworkers [309] combined the ISHE-based STE with a photoconductive switch and demonstrated a modulation of THz amplitude and the center frequency by the bias current with a significant enhancement at low THz frequencies.

In terms of operational temperature $T_0$, the choice should aim at a region in the $M(T)$ curve of the magnetic layer, that maximizes $\Delta M/\Delta T$ [113]. In other words, the pump-induced transient change in electronic temperature $\Delta T = T_{e,\text{peak}} - T_0$, with the electronic peak pump-induced temperature $T_{e,\text{peak}}$, should lead to a maximized change in magnetization $\Delta M$. Accordingly, cooling down of typical STEs did not change their performance drastically [122, 128] since typical Curie temperatures, that is, regions where $\Delta M/\Delta T$ is large, are found at around 1000 K for typical bulk 3d ferromagnets. Note however that these critical temperatures might be reduced in thin films [228] and close to interfaces [229].

Future goals include the on-chip integration of a STE [310, 311], which might boost the bandwidth of on-chip Auston switches that were recently exploited to trigger ultrafast switching of magnetic nanostructures [312] and pave the way toward THz photonic integrated circuits.

5.5 *The ideal STE: Estimating future improvements*

Finally, we would like to present an estimate of what improvements might be achievable in terms of the performance of the STE. Starting from the trilayer STEs with optimized thickness and materials (TeraSpinTec GmbH), future developments might ideally reach the following goals: (i) increase of the S2C from currently about 10% to 100%, i.e., enhance the efficiency of the ISHE or IREE S2C to fully use the initial spin current [99], (ii) utilization of 100% of the pump beam energy instead of currently about 50%, (iii) combination of forward and backward propagating THz pulses, which would yield a factor of 2 in THz-electric-field amplitude, (iv) optimized outcoupling efficiency of the THz electric field from the metal stacks into free space from currently about $Z = Z_0/(n_1 + n_2 + Z_0 G) \approx Z_0/5$ to an ideal value of $Z_0/2$ ($n_1 = n_2 = 1$, $G = 0$), implying an increase by a factor of 2.5 that might by feasible by proper photonic design [108], and (v) maximizing the injected spin current into the NM, i.e., the factor $t_{\text{FM/NM}} j_s^0$, which can be estimated from a combined spin pumping/THz emission study but has so far only been characterized for a very limited set of FM|NM combinations [226]. However, the value of $t_{\text{FM/NM}}$ has an upper limit that can be calculated from the Sharvin conductances of the materials that form the interface for spin current transmission [90, 313]. In terms of damage threshold that is currently already at 5 mJ/cm$^2$, an advanced heat-sinking strategy might help to push this limit even further [70].

In total, the above points promise an ideal increase in amplitude of the THz-electric field from the STE at a given pump-pulse energy by a factor of 100 without even considering possible improvements in the factor $t_{\text{FM/NM}} j_s^0$. This consideration implies an increase in THz pulse energy at constant pump pulse energy by 4 orders of magnitude, which demonstrates the enormous potential of future STE optimizations.

## 6. Conclusions and outlook

The field of spintronic THz emitters continues to evolve rapidly with new developments in terms of materials and of the understanding of the fundamental physics behind the STE. We see great potential to increase the emitted THz field strength by exploiting new photonic designs, emerging material classes such as topological materials and ferromagnet/semiconductor interfaces or novel spintronic phenomena such as the orbital Hall effect [314, 315].

The manifold of existing STE utilizations, for instance spintronic characterization, THz near-field imaging or THz-driven scanning tunneling microscopy, clearly demonstrates the STE's potential for both, basic research and future applications. We hope that this reviewing Editorial will be a useful introduction for the readers to the basic concepts and ideas of STEs, thereby allowing them to push the research field of spintronic THz emitters forward in the near future. Finally, the broadband and efficient detection of THz pulses using spintronic principles remains a major open research field that might significantly benefit from the knowledge acquired from studying STEs.

**Acknowledgements**

This work was supported by the National Natural Science Foundation of China (Grants Nos. 11974070, 11734006, 62027807, 12004067), the Frontier Science Project of Dongguan (2019622101004), the CAS Interdisciplinary Innovation Team, the European Union through the ERC H2020 CoG project TERAMAG/Grant No. 681917 and the German Research Foundation (DFG) through the collaborative research center SFB TRR 227 "Ultrafast spin dynamics" (project ID 328545488, projects A05 and B02) and the priority program SPP2314 ITISA (project ITISA). The authors acknowledge financial support from the Horizon 2020 Framework Programme of the European Commission under FET-Open Grant No. 863155 (s-Nebula).